\documentclass[preprint2]{aastex}

\newcommand{\myemail}{vonpapen@geo.uni-koeln.de}

\shorttitle{Forward Modeling of Power Spectra From $\mathbf{k}$-Space}
\shortauthors{von Papen and Saur}

\begin{document}

%% LaTeX will automatically break titles if they run longer than
%% one line. However, you may use \\ to force a line break if
%% you desire.

\title{Forward Modeling of Reduced Power Spectra From Three-Dimensional $\mathbf{k}$-Space}

%% Use \author, \affil, and the \and command to format
%% author and affiliation information.
%% Note, that \email has replaced the old \authoremail command
%% from AASTeX v4.0. You can use \email to mark an email address
%% anywhere in the paper, not just in the front matter.
%% As in the title, use \\ to force line breaks.

\author{Michael von Papen and Joachim Saur}
\affil{Institute of Geophysics \& Meteorology, University of Cologne, Germany}
\email{\myemail}

% \and
% 
% \author{R. J. Hanisch\altaffilmark{5}}
% \affil{Space Telescope Science Institute, Baltimore, MD 21218}

%% Notice that each of these authors has alternate affiliations, which
%% are identified by the \altaffilmark after each name.  Specify alternate
%% affiliation information with \altaffiltext, with one command per each
%% affiliation.

% \altaffiltext{1}{Visiting Astronomer, Cerro Tololo Inter-American Observatory.
% CTIO is operated by AURA, Inc.\ under contract to the National Science
% Foundation.}
% \altaffiltext{2}{Society of Fellows, Harvard University.}
% \altaffiltext{3}{present address: Center for Astrophysics,
%     60 Garden Street, Cambridge, MA 02138}
% \altaffiltext{4}{Visiting Programmer, Space Telescope Science Institute}
% \altaffiltext{5}{Patron, Alonso's Bar and Grill}

%% Mark off your abstract in the ``abstract'' environment. In the manuscript
%% style, abstract will output a Received/Accepted line after the
%% title and affiliation information. No date will appear since the author
%% does not have this information. The dates will be filled in by the
%% editorial office after submission.

\begin{abstract}
We present results from a numerical forward model to evaluate one-dimensional reduced power spectral densities (PSD) from arbitrary energy distributions in $\mathbf{k}$-space. In this model, we can separately calculate the diagonal elements of the spectral tensor for incompressible axisymmetric turbulence with vanishing helicity. Given a critically balanced turbulent cascade with $k_\|\sim k_\perp^\alpha$ and $\alpha<1$, we explore the implications on the reduced PSD as a function of frequency. The spectra are obtained under the assumption of Taylor's hypothesis. We further investigate the functional dependence of the spectral index $\kappa$ on the field-to-flow angle $\theta$ between plasma flow and background magnetic field from MHD to electron kinetic scales. We show that critically balanced turbulence {asymptotically} develops toward $\theta$-independent spectra with a slope corresponding to the perpendicular cascade. This occurs at a {transition} frequency $f_{2D}(L,\alpha,\theta)$, which is analytically estimated and depends on outer scale $L$, critical balance exponent $\alpha$ and field-to-flow angle $\theta$. We discuss anisotropic damping terms acting on the $\mathbf{k}$-space distribution of energy and their effects on the PSD. Further, we show that the spectral anisotropies $\kappa(\theta)$ as found by \citet{Horbury2008} and \citet{Chen2010a} in the solar wind are in accordance with a damped critically balanced cascade of kinetic Alfv\'en waves. We also model power spectra obtained by \citet{vonPapen2014b} in Saturn's plasma sheet and find that the change of spectral indices {inside $9\,R_\mathrm{s}$} can be explained by damping on electron scales.
\end{abstract}

%% Keywords should appear after the \end{abstract} command. The uncommented
%% example has been keyed in ApJ style. See the instructions to authors
%% for the journal to which you are submitting your paper to determine
%% what keyword punctuation is appropriate.


%% From the front matter, we move on to the body of the paper.
%% In the first two sections, notice the use of the natbib \citep
%% and \citet commands to identify citations.  The citations are
%% tied to the reference list via symbolic KEYs. The KEY corresponds
%% to the KEY in the \bibitem in the reference list below. We have
%% chosen the first three characters of the first author's name plus
%% the last two numeral of the year of publication as our KEY for
%% each reference.

%% Authors who wish to have the most important objects in their paper
%% linked in the electronic edition to a data center may do so by tagging
%% their objects with \objectname{} or \object{}.  Each macro takes the
%% object name as its required argument. The optional, square-bracket 
%% argument should be used in cases where the data center identification
%% differs from what is to be printed in the paper.  The text appearing 
%% in curly braces is what will appear in print in the published paper. 
%% If the object name is recognized by the data centers, it will be linked
%% in the electronic edition to the object data available at the data centers  
%%
%% Note, that for sources with brackets in their names, e.g. [WEG2004] 14h-090,
%% the brackets must be escaped with backslashes when used in the first
%% square-bracket argument, for instance, \object[\[WEG2004\] 14h-090]{90}).
%%  Otherwise, LaTeX will issue an error. 

\section{Introduction}
Plasma turbulence has been analyzed extensively using in-situ measurements of magnetic field and velocity fluctuations in the solar wind \citep{Matthaeus1982,Tu1984,Burlaga1986,Bruno2005}. The statistical properties of these fluctuations are usually interpreted by means of power spectral density $P(f)$ as a function of frequency in spacecraft frame. Power spectral densities (PSD) derived from these in-situ measurements of single spacecraft are generally obtained in a one-dimensional reduced form. This means that fluctuations associated with various wave vectors are observed at the same frequency in spacecraft frame and thus contribute to the spectral density $P(f)$ at a certain frequency $f$. However, turbulence models are commonly formulated in three-dimensional wave vector space, where each wave vector $\mathbf{k}$ is assigned a certain energy. Under the assumption of Taylor's hypothesis and statistical homogeneity, the observable power spectrum $P(f)$ can be obtained from the energy distribution in $\mathbf{k}$-space by a two-dimensional integration. Although this fact is very well known \citep[e.g.][]{Fredricks1976,Forman2011,Wicks2012,Turner2012b}, there is currently no application, where several plasma scales in $\mathbf{k}$-space have been considered simultaneously. We provide here a numerical tool, which we use to study the effects of various $\mathbf{k}$-space distributions.\footnote{The numerical code is made available in the supplementary online material and under \url{www.geomet.uni-koeln.de/en/research/turbulence} to be used for other $\mathbf{k}$-space distributions by the interested reader.}

Note, that for the calculation of reduced power spectral densities in frequency space, the applicability of Taylor's hypothesis and homogeneity needs to be justified for each physical system under investigation, e.g., the solar wind or magnetospheric plasmas. We will return to this issue when we compare observations in the solar wind and Saturn's magnetosphere with our model results. The transformation from $\mathbf{k}$-space to the observed frequency space taken in a rest frame moving with respect to the turbulent media is unique. In contrast, the observations taken in the moving frame do not allow to uniquely determine the spatio-temporal turbulent structure in the rest frame of the media \citep[e.g.][]{Fredricks1976}.  Thus interpretation of observed single spacecraft frequency spectra can only test whether certain $\mathbf{k}$-space distributions of turbulence are consistent with observations or not, but they cannot uniquely identify one certain $\mathbf{k}$-space structure.

The statistical properties of turbulent fluctuations have led to the understanding that turbulent fluctuations in plasmas are anisotropic with respect to a background magnetic field $\mathbf{B}_0$. \citet{Matthaeus1990} observed anisotropy with a composite slab and 2-D model consisting of wave vectors parallel and perpendicular to the mean magnetic field, respectively. \citet{Bieber1996} successfully explained the relative ratios of turbulence power in different magnetic field components of Helios measurements with a 85\% 2-D and 15\% slab model. \citet{Saur1999} found that low frequency turbulent fluctuations in the solar wind could be explained with a combination of 2-D and a radial slab model. To explain the anisotropic cascading process, \citet{Goldreich1995} proposed a critical balance between Alfv\'en wave period and nonlinear time, which leads to spectral anisotropy of the measured power spectral densities. Although there have been recent results that seem to confirm the critical balance theory \citep{Horbury2008,Podesta2009,Wicks2010,Chen2010a,TenBarge2012c,He2013}, it is an ongoing debate if critical balance is an inherent property of plasma turbulence. \citet{Tessein2009}, e.g., found no dependence of the spectral index on the field-to-flow angle $\theta$, {i.e., the angle between $\mathbf{B}_0$ and plasma flow $\mathbf{v}$ (relative to the observing spacecraft),} using a global background magnetic field in their analysis and \citet{Grappin2010} recently found a $\theta$-independent range, which they claim contradicts the critical balance assumption.

In this paper, we analyze in detail the spectral form of one-dimensional reduced PSD based on critical balance in the MHD and kinetic range of scales. The PSD are modeled for the first time for oblique angles $0^\circ\leq\theta\leq90^\circ$ from MHD to electron scales based on a three-dimensional energy distribution. In section \ref{sec:de}, we {discuss} the diagonal elements of the spectral tensor, which depend on two scalar functions for toroidal and poloidal fluctuations and we show {how} the tensor reduces to a one-dimensional spectrum if Taylor's hypothesis holds. In section \ref{sec:models}, we introduce our model for the energy distribution in $\mathbf{k}$-space based on a critically balanced cascade on MHD, ion kinetic, and electron kinetic scales. Further, we discuss potential anisotropic damping terms. In a comparative study in section \ref{sec:results}, we analyze if the results published by \citet{Horbury2008} in the MHD regime and \citet{Chen2010a} in the kinetic regime of the solar wind can be explained by a critically balanced or slab/2-D model. In {addition} to \citet{Forman2011}, who made a similar analysis for the results of \citet{Horbury2008}, we include kinetic scalings and damping effects. In section \ref{sec:satmod}, we investigate how spectra observed by \citet{vonPapen2014b} in Saturn's magnetosphere are consistent with a critically balanced cascade and damping processes.

We like to point out that our analysis is also relevant to modeling cosmic ray transport in turbulent media (e.g., Jokipii, 1966; Engelbrecht \& Burger, 2014, 2015). Cosmic rays are traveling with very large velocities through turbulent plasma fields while being scattered by time-variable fields, which are seen in a Lagrangian frame by the moving particles commonly described in an appropriate guiding center. Thus, cosmic rays are subject to the same sampling effects of anisotropic wave vector distribution as the time-dependent spacecraft measurements commonly also taken at large velocities with respect to the turbulent plasma. The effect of field-to-flow angle on cosmic ray properties has for example been studied by \citet{Bieber1996}.

\section{Diagonal Elements of the Spectral Tensor}\label{sec:de}
As we are primarily interested in the power spectra {of the fluctuations}, we will only discuss the diagonal elements of the spectral tensor. Under the assumption of incompressible axisymmetric and mirror symmetric turbulence with regards to the mean magnetic field, a general form of the correlation tensor can be derived \citep{Matthaeus1981, Montgomery1981}. Mirror symmetry leads to vanishing helicity \citep{Oughton1997} and is therefore equivalent to balanced turbulence, i.e., the same amount of energy in waves parallel and anti-parallel to the background magnetic field \citep[for the imbalanced case, see][]{Lithwick2003}. The diagonal elements of the spectral tensor then depend on only two independent scalar functions $\Psi$ and $\Phi$, which describe the toroidal (shear-Alfv\'en mode on MHD scales) and poloidal fluctuations (pseudo Alfv\'en mode), respectively. The latter are the incompressible limit of the slow mode \citep{Goldreich1995,Cho2002}.

In a magnetic field-aligned coordinate system with $\mathbf{B}_0 \parallel \mathbf{e}_z$ and $\mathbf{e}_x$ and $\mathbf{e}_y$ arbitrarily oriented perpendicular unit vectors perpendicular to $\mathbf{e}_z$, the diagonal elements can be written as \citep{Oughton1997,Wicks2012}
\begin{eqnarray}
 S_{xx}(\mathbf{k})&=&\frac{k_y^2}{k_\perp^2}\Psi + \frac{k_x^2 k_\|^2}{k_\perp^2 k^2} \Phi \label{eq:sxx} \\
 S_{yy}(\mathbf{k})&=&\frac{k_x^2}{k_\perp^2}\Psi + \frac{k_y^2 k_\|^2}{k_\perp^2 k^2} \Phi \label{eq:syy} \\
 S_{zz}(\mathbf{k})&=&\frac{k_\perp^2}{k^2} \Phi \label{eq:szz} \ ,
\end{eqnarray}
where $k_\perp^2=k_x^2+k_y^2$ {and $k_\|=k_z$}. The trace of the tensor is $\mbox{tr}\!\left(S(\mathbf{k})\right)=\Psi+\Phi$. Because the general form of $\Psi$ and $\Phi$ is not known, we need to introduce reasonable estimates for the scalar functions $\Psi$ and $\Phi$. According to \citet{Lithwick2001}, slow mode waves are passively cascaded by Alfv\'en waves, which means that the same scaling can be expected for the poloidal fluctuations $\Phi$. Due to the fact that $k_\perp\gg k_\|$, which is commonly observed in plasma turbulence \citep{Matthaeus1990,Bieber1996}, we may estimate the proportionality between the two functions from the measured {power} anisotropy
\begin{equation}
  \frac{P_\perp}{P_\|} \approx \frac{S_\perp}{S_\|} = \frac{k^2}{k_\perp^2}\frac{\Psi}{\Phi} + \frac{k_\|^2}{k_\perp^2} \approx \frac{\Psi}{\Phi} \ , \label{eq:TP}
\end{equation}
{where $P_\perp=P_{xx}+P_{yy}$ is the perpendicular and $P_\|=P_{zz}$ the parallel component of the one-dimensional reduced spectrum.} Observations in the solar wind indicate that usually $P_\perp\gg P_\|$ \citep{Hamilton2008,TenBarge2012}, which means the error {stemming from the assumption of passively cascading poloidal fluctuations} should be rather small {for the one-dimensional reduced spectra $P(f)=P_{xx}+P_{yy}+P_{zz}$}. However, we note that the influence of poloidal fluctuations increases in the kinetic range and that recent findings indicate a slightly different scaling of the poloidal fluctuations \citep{Chen2010a,Podesta2012}.

As we are able to estimate the scalar function $\Phi$ according to equation~(\ref{eq:TP}), we can now proceed to describe the diagonal elements of the spectral tensor with a single scalar function $\Psi$. This function depends on the three-dimensional wave vector $\mathbf{k}$ but since we assume axisymmetry along the mean magnetic field, this dependence further reduces to $\Psi(\mathbf{k})=\Psi(k_\perp,k_\|)$. Each wave vector represents a fluctuation with a certain wavelength along a certain direction in space. These fluctuations have a constant angle to the mean magnetic field, i.e., global and local mean magnetic field are identical in our model. For magnetic power spectral densities, the scalar function $\Psi$ is in units of nT$^2\,$m$^3$.

\subsection{Power Spectral Density in Frequency Space}
In order to compare modeled energy densities in three-dimensional $\mathbf{k}$-space (e.g., equation~(\ref{eq:emhd})) with measured PSD, we need to calculate the reduced one-dimensional spectrum in frequency space $P(f)$ from the spectral tensor given in equations (\ref{eq:sxx})-(\ref{eq:szz}). {As the modeled spectra are calculated with the assumption of statistically homogeneous fields and Taylor's hypothesis, the measured data, which the modeled spectra will be compared with, need to be statistically stationary. When our model is compared to observations, therefore both assumptions, i.e., statistical stationarity and Taylor's hypothesis need to be verified for the particular properties of the turbulent system and the relative velocity of the moving rest frame with respect to the plasma under investigation. For the application of Taylor's hypothesis, it is required that} $\mathbf{k}\cdot\mathbf{v}\gg\omega$, where $\mathbf{k}\cdot\mathbf{v}$ is the frequency of structures or fluctuations being convected over the spacecraft with velocity $\mathbf{v}$ and $\omega$ is the frequency of the corresponding wave or fluctuation in the plasma frame. {Time-stationarity of observed turbulent fields is often checked for with stationarity tests \citep[e.g.][]{Matthaeus1982}. Approximately, it can also be estimated with visual inspection of wavelet scalograms.}

{In the solar wind, where most of our subsequent analysis is applied to, Taylor's hypothesis is well satisfied in the inertial range \citep{Matthaeus1982,Narita2013,Howes2014}. However, when the Mach number drops below unity, e.g., close to the Sun or in planetary magnetospheres, Taylor's hypothesis might not be valid anymore, in general. Violation of Taylor's hypothesis may also happen at kinetic scales, where phase speeds of kinetic Alfv\'en waves (KAW) and whistler waves can become comparable to the relative plasma velocity $v$ and frequencies $\omega$ in the plasma frame strongly increase}.

{However, strong anisotropy of the fluctuations, e.g., $k_\perp\gg k_\|$, may help to satisfy Taylor's hypothesis: For KAW, the phase velocity is strongly reduced for nearly perpendicular propagation as $v_\mathrm{ph,KAW} \propto \cos(\theta)$. Therefore, Taylor's hypothesis holds in the solar wind for a critically balanced KAW cascade} even in the dissipation range \citep{Howes2014}. {The same reasoning applies to strongly anisotropic KAW fluctuations in the magnetospheres of Jupiter and Saturn \citep{Saur2002, vonPapen2014b}. Our model is also applicable for non-relativistic cosmic rays. Here, typical energies of $10\,$MeV correspond to $\sim 4\cdot 10^7\,$m$\,$s$^{-1}$, for which Taylor's hypothesis is clearly satisfied. Whistler wave turbulence on the other hand violates Taylor's hypothesis and can therefore not be described by our model.}

If Taylor's hypothesis and homogeneity apply, the diagonal elements of the reduced spectral tensor in frequency space, $P_{ii}(f)$, can be obtained by integrating the three-dimensional energy densities $S_{ii}(\mathbf{k})$ over a plane perpendicular to the flow direction $\mathbf{v}$ \citep{Fredricks1976}:
\begin{eqnarray}
 P_{ii}(f)&=&\!\!\!\!\!\!\!\!\int\limits_{-\infty}^\infty \!\!\mathrm{d}t \, \mathrm{e}^{i 2\pi f t} \int\limits_{-\infty}^\infty \!\mathrm{d}k^3 \, S_{ii}(\mathbf{k}) \, e^{-i \mathbf{k}\cdot\mathbf{v}t} \nonumber \\
 \phantom{P_{ii}(f)} &=&\!\!\!\!\!\!\!\! \int\limits_{-\infty}^\infty \!\!\mathrm{d}^3\!k \, S_{ii}(\mathbf{k}) \, \delta\!\left(2\pi f{-}k_x v\sin(\theta){-}k_z v\cos(\theta)\right) . \quad \label{eq:trafo}
\end{eqnarray}
Here, we assumed that the $z$-axis is parallel to the background magnetic field ($\mathbf{e}_z \parallel \mathbf{B}_0$) and ${\mathbf{e}_y=\frac{\mathbf{e}_z \times \mathbf{v}}{|\mathbf{e}_z \times \mathbf{v}|} }$, so that $\mathbf{e}_x$ lies in the plane spanned by $\mathbf{B}_0$ and $\mathbf{v}$, which is sometimes called the quasi-parallel direction \citep{Bieber1996}.

In order to evaluate equation~(\ref{eq:trafo}) numerically, we apply a coordinate transformation that reduces the dimensions of the integration. If we rotate the coordinate system by the angle $90^\circ-\theta$ about the $y$-axis such that 
\begin{eqnarray}
	k_x' &=&\phantom{-}k_x \sin(\theta) +k_z \cos(\theta) \\
	k_y' &=&\phantom{-}k_y \\
	k_z' &=&-k_x \cos(\theta) + k_z \sin(\theta) \ ,
\end{eqnarray}
then $k_x'$ will be aligned with $\mathbf{v}$ and the plane of integration lies in the $y'{-}z'$ plane. {The unprimed parallel and perpendicular wave numbers now take the form}
\begin{eqnarray}
	 k_\perp &=& { \sqrt{[k_x' \sin(\theta) -k_z' \cos(\theta)]^2 + k_y'^2}} \\
	 k_\| &=& { k_x' \cos(\theta) + k_z' \sin(\theta)} \ .
\end{eqnarray}
In the primed coordinate system equation~(\ref{eq:trafo}) becomes
\begin{equation}
 P_{ii}(f)= \frac{1}{v}\int_{-\infty}^\infty \!\mathrm{d}^3\!k' \, S_{ii}(\mathbf{k}') \, \delta\!\left( \frac{2\pi f}{v}{-}k_x'\right) \ . \label{eq:pfull}
\end{equation}
Now the Delta function can be evaluated and the integration over three dimensions in equation~(\ref{eq:trafo}) is thus reduced to two dimensions $k_y'$ and $k_z'$, which is much easier to {calculate} numerically. Note, that several different wave vectors map to the same frequency in spacecraft frame. In fact, the power spectral density $P_{ii}(f)$ is determined by integration over a plane with normal vector $\mathbf{k}_\mathrm{n}=\left(k_x \sin(\theta),0,k_z \cos(\theta)\right)$. For the numerical evaluation, the scalar functions $\Psi'$ and $\Phi'$ are inserted in equation~(\ref{eq:pfull}) using equations (\ref{eq:sxx})-(\ref{eq:szz}) in the rotated form. With this, we are able to calculate the power spectral densities $P_{ii}(f)$ in spacecraft frame.

\section{Results for a Critically Balanced Turbulence Model} \label{sec:models}
In this section, we introduce a $\mathbf{k}$-space turbulence model ranging from MHD to electron scales based on commonly discussed models in the literature. With this model we calculate one-dimensional reduced PSD for arbitrary measurement geometries. We focus on the theory of a strong critically balanced turbulent cascade as originally proposed by \citet{Goldreich1995} for Alfv\'enic turbulence. It has been further developed to include the ion kinetic range \citep{Howes2008,Schekochihin2009} and the electron kinetic range \citep{TenBarge2013}. Observations indicate that it is suited to describe solar wind turbulence \citep{Horbury2008,Wicks2010,Chen2010a}, although these observations have not been explicitly compared with reduced spectra rigorously derived from three-dimensional $\mathbf{k}$-space models. However, with the forward model derived here, we are able to perform such an explicit test and are able to check if the observations are truly in agreement with critical balance and other suggested properties of three-dimensional $\mathbf{k}$-space.

\subsection{Critically Balanced Cascade on MHD Scales}
According to \citet{Goldreich1995}, the (kinetic) energy distribution on MHD scales can be described as
\begin{equation}
 E(\mathbf{k})\sim\frac{V_\mathrm{A}^2}{k_\perp^{10/3} L^{1/3}} \ f\!\left(L^{1/3}\frac{k_\|}{k_\perp^{2/3}}\right) \ ,\label{eq:gs95}
\end{equation}
where ${ V_\mathrm{A}=B_0/\sqrt{\mu_0 \varrho} }$ is the Alfv\'en speed with $\varrho$ the mass density of the plasma, $L$ the outer scale, and ``\textit{$f(u)$ is a positive, symmetric function of $u$, that becomes negligibly small, when $|u|\gg1$}'' \citep{Goldreich1995}. It is this function $f(u)$, which contains the critical balance relation of wave vectors on MHD scales
\begin{equation}
 k_\| \sim L^{-1/3} k_\perp^{2/3} \label{eq:cb4} \ .
\end{equation}
The energy is supposed to be isotropically injected into the system on the outer scale $L$, where the excitation is assumed to be strong $\delta v_L\sim V_\mathrm{A}$. This relation leads to spectral anisotropy, i.e., different scalings depending on the field-to-flow angle $\theta$. \citet{Cho2002} find that the function that best explains the magnetic fluctuations in their simulation of incompressible MHD turbulence, is
\begin{equation}
 E_\mathrm{MHD}=\left(\frac{B_0^2}{L^{1/3}}\right) k_\perp^{-10/3} \exp\left({-}L^{1/3}\frac{|k_\||}{k_\perp^{2/3}}\right) \label{eq:emhd} \ .
\end{equation}
According to equation~(\ref{eq:gs95}), this indicates that the magnetic fluctuations on the outer scale $L$ are strong, $\delta B\sim B_0$. However, \citet{Forman2011} find from fits to solar wind data that $B_0^2$ overestimates the energy of the turbulent fluctuations by a factor of $2-3$. Therefore, the energy level of the modeled power spectra cannot be obtained unambiguously. The energy distribution given by equation~(\ref{eq:emhd}) describes the Alfv\'enic fluctuations and we thus set $\Psi=E_\mathrm{MHD}$ on MHD scales. According to equation~(\ref{eq:TP}), this constrains the poloidal function to $\Phi=\frac{P_\|}{P_\perp}\Psi$. 

Next to the exponential function used in equation~(\ref{eq:emhd}), there are several different possibilities to describe $f(u)$, e.g., Dirac Delta, Heaviside or Gauss functions. The particular choice of $f(u)$, however, results in similar spectral anisotropies as found in the case of MHD scales by \citet{Forman2011} and \citet{Turner2012b}. For $\theta=90^\circ$, the one-dimensional PSD scales as $P(f)\propto f^{-5/3}$ and for $\theta=0^\circ$ as $P(f)\propto f^{-2}$ in the inertial range. Exact power-laws in frequency space only occur for these two extreme cases. In the literature, however, it is often implicitly assumed that the spectra in the intermediate range $0^\circ<\theta<90^\circ$ also follow power-laws with spectral slopes $5/3<\kappa<2$ and that the spectral index $\kappa$ of the reduced spectra is {-- similar to the spectral index of three-dimensional $\mathbf{k}$-space --} independent of frequency. In this paper, we will see that these assumptions are incorrect.

\subsection{Transition From MHD to Kinetic Scales}\label{sec:kaw}
When the observed Doppler-shifted frequencies reach scales close to the characteristic ion scales, such as the ion gyro radius $\rho_i$ or the ion inertial length $\lambda_i$, the MHD approximation breaks down and one needs to take into account kinetic effects to describe the plasma dynamics. For the remainder of this paper, we use the gyro radius as the controlling kinetic scale with an associated critical balance on ion kinetic scales
\begin{equation}
  k_\|\sim L^{-1/3} \rho_i^{-1/3} k_\perp^{1/3} \label{eq:cb_kaw2}
\end{equation}
{according to \citet{Howes2008}. Here, we drop the factor $\left( \beta_i +2/(1+T_e/T_i)\right)^{1/6}$, where $\beta_i$ is the ion plasma beta and $T_i$ and $T_e$ the temperatures of ions and electrons, respectively, because the factor is of order unity in the solar wind and Saturn's magnetosphere.}  {Note, that our forward model can similarly be formulated for other theories or with the inertial length as the controlling scale.}

We model the transition from MHD to kinetic scales as an abrupt change in the function $E_\mathrm{MHD}(\mathbf{k})$ at $k_\perp \rho_i=1$ without energy loss, e.g., due to ion resonances. Let us denote the function $E_\mathrm{MHD}$, given in equation~(\ref{eq:emhd}), as the one applicable for $k_\perp \rho_i\leq1$ and $E_\mathrm{KAW}$ as the one applicable for $k_\perp \rho_i\geq1$. At $k_\perp \rho_i=1$, both energy distributions are equal, $E_\mathrm{MHD}(k_\perp \rho_i{=}1)=E_\mathrm{KAW}(k_\perp \rho_i{=}1)$. Also, the kinetic energy distribution, $E_\mathrm{KAW}$, is assumed to scale as $P(f)\propto f_\perp^{-7/3}$ for $\theta=90^\circ$ \citep{Howes2008}. These requirements are fulfilled in the expression
\begin{equation}
  E_\mathrm{KAW} = \left(\frac{B_0^2}{L^{1/3}\rho_i^{1/3}}\right) k_\perp^{-11/3} \exp\left({-}L^{1/3}\rho_i^{1/3}\frac{|k_z|}{k_\perp^{1/3}}\right) \label{eq:ekaw} \ .
\end{equation}
{For the parallel cascade ($\theta=0^\circ$), this leads to a scaling of $P(f)\propto f^{-5}$.}

At even smaller scales, we approach the electron gyro radius $\rho_e$ and the energy distribution changes once again from $E_\mathrm{KAW}$ to an electron or dissipation range distribution $E_\mathrm{ED}$. Here, Landau damping is assumed to weaken the cascade so that there is no more parallel transfer of energy \citep{Sridhar1994,Howes2008}. The electron or dissipation range fluctuations are modeled with an associated critical balance \citep{TenBarge2013} of
\begin{equation}
   k_\|\sim L^{-1/3} \rho_i^{-1/3}\rho_e^{-1/3}\ . \label{eq:cbed}
\end{equation}
A functional form that satisfies the equality of $E_\mathrm{KAW}$ and $E_\mathrm{ED}$ at $k_\perp \rho_e =1$ can be given by
\begin{equation}
  E_\mathrm{ED} = \left(\frac{B_0^2}{L^{1/3}\rho_i^{1/3}}\right) k_\perp^{-11/3} \exp\left({-}L^{1/3}\rho_i^{1/3}\rho_e^{1/3} |k_z|\right) \label{eq:ed} \ .
\end{equation}
It can be shown that this leads to a scaling of $P(f) \propto f^{-8/3}$ for the perpendicular cascade ($\theta=90^\circ$) and an exponential decay $P(f) \propto \exp(-f)$ for the parallel cascade.

\begin{figure}[t]
 \plotone{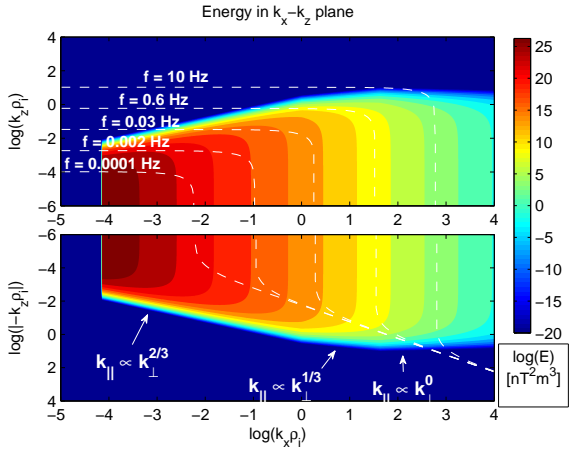}
 \caption{Logarithmically equidistant color-coding of energy {density} in $\mathbf{k}$-space calculated with equations~(\ref{eq:emhd}), (\ref{eq:ekaw}), {and (\ref{eq:ed})}. Thin dashed white lines show integration planes after equation~(\ref{eq:plane}) for logarithmically equidistant frequencies $f=10^{-4}{-}10\,$Hz {and a field-to-flow angle} $\theta=1^\circ$. Due to the double-logarithmic plot these planes appear as curved lines. {Note, that energy resides only in wave vectors $k\geq L^{-1}$.} \label{fig:kspa}}
\end{figure}

The energy distribution in $\mathbf{k}$-space according to our model is shown in Figure \ref{fig:kspa} for the parameters given in Table \ref{tab:sw}. It shows logarithmically equidistant iso-contours of energy densities according to equations (\ref{eq:emhd}), (\ref{eq:ekaw}) and (\ref{eq:ed}) in a double-logarithmic plot as a function of $k_x$ and $k_z$ normalized by the ion gyro radius $\rho_i$. The characteristic slopes of the critical balance relations, $2/3$, $1/3$, and $0$ in the MHD, KAW, and ED regime, respectively, can be seen as the boundary, where the energy becomes negligible (dark blue). Note, that the energy density decreases with $k_\|$ and $k_\perp$, which leads to a seemingly negative slope in the electron dissipation range.

Recall from equation~(\ref{eq:trafo}) and subsequent discussion that the reduced one-dimensional spectrum is calculated by integrating over a plane given by
\begin{equation}
	 k_x \sin(\theta) = \frac{2\pi f}{v} - k_z \cos(\theta)  \ . \label{eq:plane}
\end{equation}
Projections of these planes into the $k_x{-}k_z$ plane are shown as dashed lines in Figure \ref{fig:kspa} for an angle $\theta=1^\circ$ at logarithmically equidistant frequencies $f=10^{-4}{-}10\,$Hz. For $\theta=0^\circ$, these dashed lines would be horizontal and parallel to the $k_x$-axis, and for $\theta=90^\circ$, they would be vertical and parallel to the $k_z$-axis. However, due to the double-logarithmic nature of the plot, the dashed lines for intermediate angles $\theta$ are curved. Note, that even for a very small angle $\theta=1^\circ$, the plane of integration turns quasi-perpendicular to the $k_x$-axis, i.e., lies parallel to planes expected from integration for $\theta= 90^\circ$, for sufficiently large $k_x$.

\subsection{Cascade Toward Quasi-Perpendicular Spectra}\label{sec:2d}
\begin{figure}[t]
 \plotone{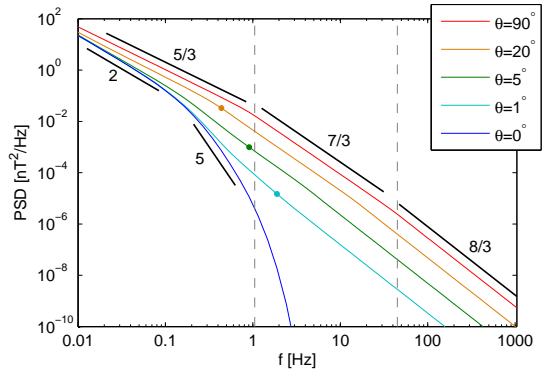}
 \caption{PSD for different angles $\theta$ as a function of frequency $f$ with plasma parameters given in Table \ref{tab:sw}. The underlying critically balanced $\mathbf{k}$-space structure consists of three different regimes with spectral indices $5/3$, $7/3$ and $8/3$ for $\theta=90^\circ$. Note, that no damping is included. Vertical dashed lines show Doppler-shifted gyro-radii $\rho_i$ and $\rho_e$ for $\theta=90^\circ$. At high frequencies all spectra tend toward a slope of $7/3$ in KAW and $8/3$ in ED range. Colored dots show the analytically approximated transition frequency $f_{2D}^\mathrm{a}$, where the slopes of the spectra approach the slope of the perpendicular spectrum. $f_{2D}^\mathrm{a}$ is calculated with equation~(\ref{eq:fmax}) assuming a pure KAW cascade. \label{fig:psd_theta}}
\end{figure}

While equation~(\ref{eq:pfull}) can be integrated analytically for angles $\theta=0^\circ$ and $\theta=90^\circ$, it can only be numerically evaluated for intermediate angles $0^\circ<\theta<90^\circ$. Here, we carry out this numerical integration for the first time considering all scales from the MHD to the {electron} kinetic regime. Results of this calculation using the parameters given in Table \ref{tab:sw} are shown in Figure \ref{fig:psd_theta}. We plot the resulting PSD for several field-to-flow angles as a function of frequency $f$ in spacecraft frame. Note, that frequency can be transformed into normalized wave number according to $\frac{2\pi f}{v}\rho_i=k\rho_i$, particularly for parameters in Table \ref{tab:sw}: $f\approx0.95\cdot k\rho_i$.

For $\theta=90^\circ$ the spectral breaks of the PSD, which mark the transition from MHD to KAW and KAW to ED range, are located at the Doppler-shifted frequencies corresponding to $\rho_i$ and $\rho_e$ (denoted by vertical dashed lines). Here, the spectral slope of the reduced PSD steepens from $5/3$ to $7/3$ and from $7/3$ to $8/3$, respectively. For smaller angles $\theta<90^\circ$ these spectral breaks are less pronounced and occur at lower frequencies. The expected spectral slopes for $\theta=0^\circ$ and $\theta=90^\circ$ are shown as black lines in Figure \ref{fig:psd_theta} to guide the eye. Here and throughout this paper, we use plasma parameters characteristic of the solar wind at $1\,$AU as given in Table \ref{tab:sw}, which we adopt from the work of \citet{Alexandrova2009}, \citet{Schekochihin2009}, and \citet{Chen2010a}.

\begin{table}[ht]
  \begin{center}
  \caption{Solar wind parameters used for our model. \label{tab:sw}}
  \begin{tabular}{l c}
    \tableline\tableline
    Parameter  &  Value  \\
    \tableline
    Background magnetic field $B_0$  & $5\,$nT   \\
    Outer scale $L$ & $10^6\,$km   \\
    Plasma speed $v$ & $600\,$km$\,$s$^{-1}$   \\
    Alfv\'en speed $V_\mathrm{A}$  & $60\,$km$\,$s$^{-1}$   \\
    Ion gyro radius $\rho_i$ & $100\,$km   \\
    Electron gyro radius $\rho_e$ & $2.3\,$km   \\
    \tableline
  \end{tabular}
  \end{center}
\end{table}

We note a particular interesting feature in Figure \ref{fig:psd_theta}: the spectra for oblique angles are steepened in a short frequency range around the first spectral break at $2\pi f/(v\sin(\theta))\rho_i=k_\perp\rho_i\sim1$ and then flatten out toward a slope of $7/3$ at higher frequencies. The smaller the angle $\theta$, the longer the range, where PSD are steeper than $7/3$. Such an asymptotic behavior has been predicted for the MHD range by \citet{Forman2011}, however, without specifying the frequency range, where this transition could be expected. Here, we show for the first time that this transition happens on a short frequency range around the spectral break in the solar wind. We emphasize the importance of this result as it contradicts the commonly assumed constancy of the spectral index with frequency. Also, it implies that PSD in the kinetic range are almost exclusively observed with their perpendicular slope (here $7/3$ and $8/3$) if not subject to damping. In the remainder of this section, we explain this feature as a geometrical or sampling effect.

The flattening of the PSD, although puzzling at first, can be explained with the anisotropic distribution of energy in $\mathbf{k}$-space. Because of the linear relation between $k_x$ and $k_z$ in equation~(\ref{eq:plane}), the point of maximum curvature {of the integration plane in Figure~\ref{fig:kspa}} grows faster than the critical balance relations $k_\|\sim k_\perp^\alpha$ in the MHD ($\alpha=2/3$), {KAW} ($\alpha=1/3$), and ED ($\alpha=0$) range. Hence, for increasing frequencies $f$, the plane of integration given by equation~(\ref{eq:plane}) effectively has only contributions from the part {quasi-}perpendicular to the $k_x$-axis in the double-logarithmic plot of Figure \ref{fig:kspa}. Therefore, at frequencies above a transition frequency $f_{2D}$, the power spectrum at a given field-to-flow angle $\theta$ will have a slope within a small error $\Delta\kappa$ to that at $\theta=90^\circ$ appropriate to MHD ($5/3$), KAW ($7/3$), or ED ($8/3$). The transition frequency $f_{2D}$ thus marks the boundary, where the anisotropy of the $\mathbf{k}$-space energy distribution becomes so large that it is indistinguishable from a 2D-distribution in the reduced PSD within measurement error $\Delta\kappa$. It is this feature that causes the spectral slope of the PSD to asymptotically approach its perpendicular value at high frequencies.

A derivation of an analytically approximated expression $f_{2D}^\mathrm{a}$ (see equation~(\ref{eq:fmax})) can be found in Appendix \ref{app}. The frequencies, where the PSD turn quasi-perpendicular, are shown in Figure \ref{fig:psd_theta} as colored dots assuming a pure KAW cascade. Note, that the transition frequency $f_{2D}^\mathrm{a}$, in contrast to the spectral break, shifts to higher frequencies for smaller angles $\theta$. Assuming a plain MHD cascade, the approximate transition frequency from equation~(\ref{eq:fmax}) gives $f_{2D,\mathrm{MHD}}^a(\theta{=}20^\circ)=0.2\,$Hz. The corresponding transition frequencies for $\theta=[1^\circ,5^\circ]$ reach into the kinetic range of scales. The transition to quasi-perpendicular scaling is always found at frequencies below the electron spectral break, $k_\perp\rho_e\sim1$, which is why frequencies $f_{2D}^\mathrm{a}$ are not shown in the ED range.

We find that the transition for KAW toward a quasi-perpendicular cascade for almost all field-to-flow angles (namely $\theta>4^\circ$, see Figure \ref{fig:fmax} in the appendix) occurs already below $1\,$Hz, where the spectral break is observed at typical solar wind conditions \citep{Sahraoui2010, Alexandrova2012}. We conclude that spectra of a critically balanced KAW cascade without damping on kinetic scales will almost exclusively be observed with spectral indices close to $7/3$ and $8/3$ in the ion and electron kinetic range, respectively. Only for small angles and in a short frequency range between the first spectral break and $1\,$Hz a measurably steeper slope might be observed in the solar wind.

We stress the importance of this result: we have shown that the spectral index in a critically balanced cascade for oblique field-to-flow angles $0^\circ<\theta<90^\circ$ is not constant with growing frequency. Instead, it evolves toward a quasi-perpendicular slope as a function of frequency. Therefore, we cannot expect to observe significantly steeper slopes than $7/3$ {and $8/3$} over a broad frequency range in the KAW and ED range, respectively, for a critically balanced cascade {in the solar wind}. Indeed, we see in Section~\ref{sec:damp} that such slopes only appear if the fluctuations are additionally subject to damping. This transition of reduced PSD toward quasi-perpendicular spectra contrasts a common understanding in the literature, where it is often implicitly assumed that the PSD have a power-law shape with a constant spectral index that only depends on the field-to-flow angle. It is interesting to note that a three-dimensional direct numerical simulation with strong guide field by \citet{Grappin2010} produced a qualitatively similar transition toward a perpendicular cascade with a $\theta$-independent slope, although the flattening of their spectra does not follow equation~(\ref{eq:fmax}).

\subsection{Anisotropic Damping}\label{sec:damp}
Several mechanisms have been proposed to explain the dissipation of fluctuations on small scales, e.g., ion-cyclotron damping, Landau damping and current sheet formation \citep{Matthaeus1990,Leamon1999,Dmitruk2004,Howes2009,Schekochihin2009,TenBarge2013}. Most of the proposed mechanisms are anisotropic with regards to the background magnetic field. In this section, we analyze how possible anisotropic damping terms that act on the energy distribution in three-dimensional $\mathbf{k}$-space change the characteristics of reduced one-dimensional spectra. Here, we assume that damping leads to an exponential decay of the energy in $\mathbf{k}$-space associated with the damped wave vectors.

From our assumption of critical balance, it follows that $k_\perp \gg k_\|$. Therefore, damping at a fixed scale affects the turbulent cascade first through perpendicular wave vectors and only later through wave vectors parallel to the mean magnetic field. It has been argued that the ion-cyclotron resonance at $k_\|\sim \Omega_\mathrm{ic}/V_\mathrm{A}$, {where $\Omega_\mathrm{ic}=\frac{eB}{m_i}$ is the ion-cyclotron frequency}, has only minor influence on a critically balanced cascade as it is reached only at very high perpendicular wave numbers $k_\perp$, where Landau damping already dominates \citep{Howes2008,Schekochihin2009,Cranmer2012}. With our synthetic spectra, we are now able to test this argument quantitatively.

\begin{figure}[t]
 \plotone{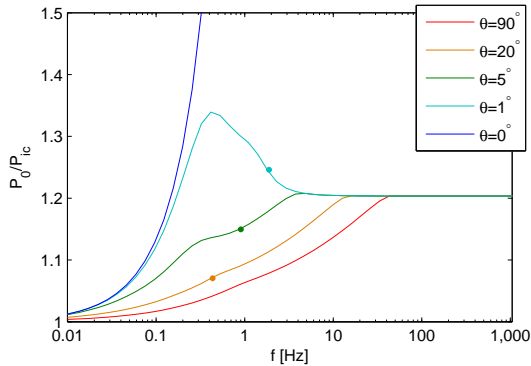}
 \caption{Ratio of PSD without damping ($P_0$) and with ion-cyclotron damping ($P_\mathrm{ic}$) according to equation~(\ref{eq:dampic}) as a function of frequency $f$ for several field-to-flow angles $\theta$. Colored dots show $f_{2D}^\mathrm{a}$ according to a pure KAW cascade. \label{fig:icr_diff}}
\end{figure}

\subsubsection{Damping of Parallel Wave Vectors}
For the ion-cyclotron resonance, the damping rate is low for $k_\|V_\mathrm{A}\ll \Omega_\mathrm{ic}$ and increases rapidly as $k_\| V_\mathrm{A} /\Omega_\mathrm{ic}$ approaches unity \citep{Howes2008,Cranmer2012}. We may therefore model the ion-cyclotron resonance as an exponential decay according to
\begin{equation}
 E_\mathrm{ic}(\mathbf{k}) = E(\mathbf{k}) \cdot \exp\left({-}\frac{k_\|V_\mathrm{A}}{\Omega_\mathrm{ic}}\right) \label{eq:dampic} \ .
\end{equation}
To quantify the effective damping, we model {damped} spectra $P_\mathrm{ic}$ using the parameters in Table \ref{tab:sw} and find only minor changes compared to their undamped counterparts $P_0$.

In Figure~\ref{fig:icr_diff}, we show the ratio $P_0/P_\mathrm{ic}$ as a function of normalized wave number for several angles $\theta$. {Both $P_0$ and $P_\mathrm{ic}$ cover all three ranges from MHD to KAW to electron kinetic scales.} For quasi-parallel spectra ($\theta=1^\circ$), a peak factor of $1.34$ is observed around the spectral break. Power spectra $P_\mathrm{ic}$ for larger field-to-flow angles $\theta\geq5^\circ$ only differ a factor of $1.2$ from PSD without damping. On electron scales, the difference between damped and undamped spectra remains constant as there is no more parallel transfer along the cascade. The asymptotic factor $P_0/P_\mathrm{ic}$ for $2\pi f/V\geq \rho_e^{-1}$ and $\theta=90^\circ$ can be calculated by integration of equation~(\ref{eq:ed}) over $k_z$ according to
\begin{eqnarray}
	&&\!\!\!\!\!\!\!\!\!\!\!\!\!\!\!\!\!\!\!\!\!\frac{P_0}{P_\mathrm{ic}}(2\pi f/V{\geq}\rho_e^{-1},\theta{=}90^\circ) \nonumber \\
	&=& \frac{\int\limits_{-\infty}^\infty \!\!\mathrm{d}k_z \exp\left({-}L^{1/3}\rho_i^{1/3}\rho_e^{1/3} |k_z|\right)}{\int\limits_{-\infty}^\infty \!\!\mathrm{d}k_z \exp\left({-}L^{1/3}\rho_i^{1/3}\rho_e^{1/3} |k_z|-|k_z|\frac{V_\mathrm{A}}{\Omega_\mathrm{ic}}\right)} \nonumber \\
	&\approx& 1.21 \ .
\end{eqnarray}
This shows that the influence of cyclotron damping on the spectral shape can be neglected in the solar wind for critically balanced turbulence.

However, one might think of systems, in which the ion-cyclotron resonance \textit{does} change the form of the PSD significantly. Such systems could be characterized by (1) low plasma density, which leads to large Alfv\'en speeds; (2) a critical balance exponent $\alpha$ close to unity; and/or (3) a much smaller outer scale than the $L\sim10^9{-}10^{10}\,$m found in the solar wind \citep{Howes2008,Schekochihin2009}, so that high $k_\|$ values are reached earlier in the cascade. An enhanced ion-cyclotron damping then leads to a visible decrease of spectral power around the spectral break.

\subsubsection{Damping of Perpendicular Wave Vectors}
\begin{figure}[t]
 \plotone{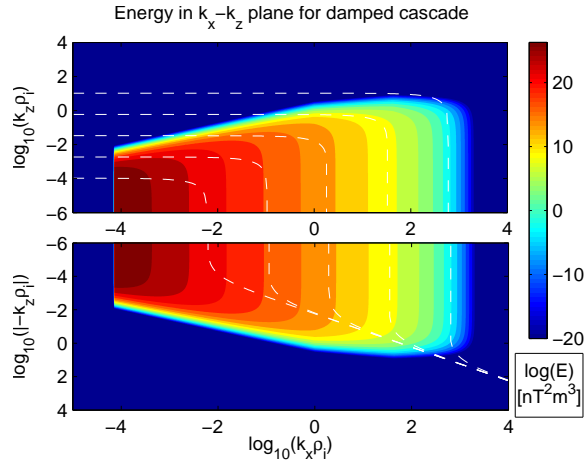}
 \caption{Exponentially damped and critically balanced energy distribution in $\mathbf{k}$-space according to equations~(\ref{eq:dampic}) and (\ref{eq:damp}). The distribution contains three ranges for MHD, ion and electron kinetic scales. Thin dashed white lines show integration planes after equation~(\ref{eq:plane}) for logarithmically equidistant frequencies $f=10^{-4}{-}10\,$Hz and a field-to-flow angle $\theta=1^\circ$. \label{fig:kspa_damp}}
\end{figure}

Let us now turn to damping of perpendicular wave vectors. \citet{Alexandrova2012} found that reduced power spectra measured in the solar wind can empirically be described by
\begin{equation}
	P(k_\perp)\propto k_\perp^{-\kappa} \exp(-k_\perp\rho_e) \label{eq:damp1D}
\end{equation}
with a spectral index of $\kappa=8/3$. This result is consistent with numerical gyrokinetic simulations \citep{Howes2011a,TenBarge2013} and we may treat the exponential damping term of equation~(\ref{eq:damp1D}) as a proxy for electron Landau damping. Note, that we apply the damping term of equation~(\ref{eq:damp1D}) to three-dimensional $\mathbf{k}$-space according to
\begin{equation}
 E_\mathrm{damp}(\mathbf{k}) = E(\mathbf{k}) \cdot \exp({-}k_\perp\rho_e) \label{eq:damp} \ .
\end{equation}
{The corresponding distribution in $\mathbf{k}$-space is shown in Figure \ref{fig:kspa_damp}, where both damping of parallel and perpendicular wave vectors (equations (\ref{eq:dampic}) and (\ref{eq:damp})) is included.} While the power at a certain frequency is obtained by integration over a two-dimensional plane, the damping term depends on $k_\perp$ and is therefore axisymmetric with respect to $k_z$. This means that the integration along $k_y$ includes perpendicular wave numbers larger than $k_\perp=2\pi f/(v\sin(\theta))$. Consequently, the effect of the exponential damping term is stronger when applied to three-dimensional $\mathbf{k}$-space compared to a reduced one-dimensional spectrum. Therefore, we are able to produce similar results to those of equation~(\ref{eq:damp1D}) { with $\kappa=8/3$}, although we apply a less steep spectral index of only $7/3$ in the ion kinetic range (see Figure \ref{fig:psd_theta_damp}).

\subsubsection{Reduced PSD Subject to Damping}
\begin{figure}[t]
 \plotone{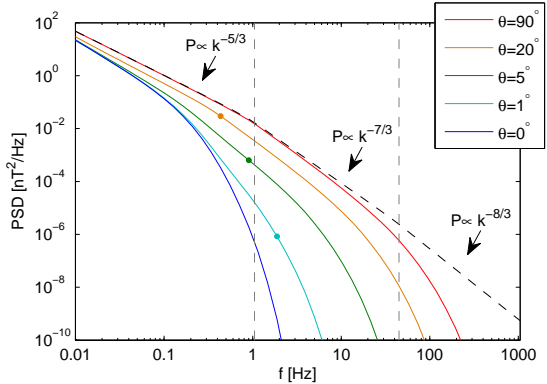}
 \caption{PSD subject to damping according to equations~{(\ref{eq:dampic}) and} (\ref{eq:damp}) as a function of frequency $f$ for parameters given in Table \ref{tab:sw}. The black dashed line shows the PSD for $\theta=90^\circ$ without damping. Colored dots show for orientation at what frequency the spectrum would turn quasi-perpendicular for a pure KAW cascade without damping. \label{fig:psd_theta_damp}}
\end{figure}

Figure \ref{fig:psd_theta_damp} shows damped PSD according to equations~{(\ref{eq:dampic}) and} (\ref{eq:damp}) with the same plasma parameters (Table \ref{tab:sw}) as in Figure \ref{fig:psd_theta}. As expected, the high frequency part is strongly affected and the damping leads to a characteristic exponential decay. The {spectra are dominated by damping of perpendicular wave vectors and the} damping terms are found to be generally more effective at small angles $\theta$. The latter can be understood considering the plane of integration given by equation~(\ref{eq:plane}). The power at a certain frequency primarily stems from perpendicular wave numbers $k_\perp$. For small angles $\theta$, this involves larger $k_\perp$, which are more strongly damped than a corresponding spectrum with angles close to $90^\circ$.

Although strong damping sets in not before electron scales in wave vector space, the KAW range of the PSD, $\rho_i^{-1}<k<\rho_e^{-1}$, is already affected: Figure \ref{fig:psd_theta_damp} shows the undamped spectrum for $\theta=90^\circ$ as black dashed line and it is fairly visible that the spectrum subject to damping is steeper. In fact, we measure a spectral slope of $2.63$ (shown by black dashed line) in the ion kinetic range, $1<k\rho_i<42$ for $P_\mathrm{damp}(\theta{=}90^\circ)$, although the corresponding energy in $\mathbf{k}$-space scales with $7/3$.

If $T_i/T_e$ decreases or the electron gyro radius increases, the exponential decay moves to lower frequencies and further steepens the spectra. The two opposed mechanisms, flattening toward perpendicular slope and damping, can cancel each other and {may} lead to a range of seemingly constant slope. This shows that an approximate power-law of the measured PSD is not equivalent to the absence of damping. Further, it shows that the observed spectral slope of the reduced spectrum on ion kinetic scales is not necessarily the spectral index predicted for this range by the underlying theory.

\section{Application to Solar Wind Observations} \label{sec:results}
By numerically evaluating equation~(\ref{eq:pfull}), it is possible to calculate power spectral densities $P_{ii}(f)$ for any component $i=x,y,z$ and given plasma parameters (plasma speed, outer scale, field-to-flow angle, background magnetic field, Alfv\'en speed, gyro radii). Here, we compare our results to in-situ measurements in the solar wind made by \citet{Horbury2008} in the MHD range and \citet{Chen2010a} in the kinetic range. \citet{Podesta2009} and \citet{Wicks2010} have presented similar results in the MHD range, which are consistent with those from \citet{Horbury2008}.
We show that the measured spectral anisotropies are in accordance with our model and can thus be described by a critically balanced cascade. However, the influence of damping turns out to be more important than previously thought.

\subsection{MHD Turbulence}\label{sec:horb}
\begin{figure}[t]
 \plotone{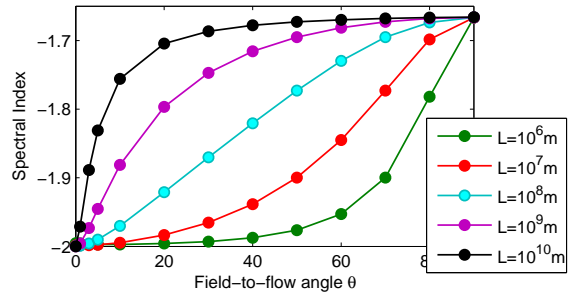}
 \caption{Spectral anisotropy in the inertial range for different values of the outer scale $L$ for MHD scaling, i.e., integration of (\ref{eq:pfull}) over expression (\ref{eq:emhd}). The spectral index is obtained in frequency range $f=[10^{-4}{-}0.1]\,$Hz. A smaller outer scale results in a transition of the spectral index from $2$ to $5/3$ at larger angles $\theta$. \label{fig:L}}
\end{figure}

From the assumption of critical balance it follows that the reduced PSD scales as $P\propto f^{-5/3}$ for $\theta=90^\circ$ and $P\propto f^{-2}$ for $\theta=0^\circ$. Figure \ref{fig:L} shows exemplary for the case of MHD, i.e., equation~(\ref{eq:emhd}) only, that the transition between these two scalings is controlled by the outer scale $L$. Here, we have calculated the spectral slope from a least-squares fit in the frequency range $f=[10^{-4},0.1]\,$Hz. Note, that the spectral index slightly varies in this frequency range and, therefore, no {exact} power-law is observed. However, the result shown in Figure \ref{fig:L} impressibly characterizes the influence of the outer scale on the cascade. It can be seen that the perpendicular scaling is {reached only at very large angles $\theta$} for small outer scales $L$. This reflects the evolution of the turbulent cascade along the critical balance path. While the energy is isotropic at the outer scale, it grows increasingly anisotropic as it cascades to smaller scales.

We can use this functional dependence of the spectral anisotropy to check whether our critically balanced $\mathbf{k}$-space distribution generates results in agreement with \citet{Horbury2008} and, if so, which value of $L$ fits the observations best. For several outer scales $L$, we calculate the spectral anisotropy and fit the spectral indices $\kappa(L,\theta)$ to their results. To evaluate the critically balanced power spectra, we use the complete set of energy distributions, $E_\mathrm{MHD}$, $E_\mathrm{KAW}$ and $E_\mathrm{ED}$, together with anisotropic damping, i.e., exponential damping at the ion-cyclotron resonance and at electron scales according to equations (\ref{eq:dampic}) and (\ref{eq:damp}), respectively.

\begin{figure}[t]
 \plotone{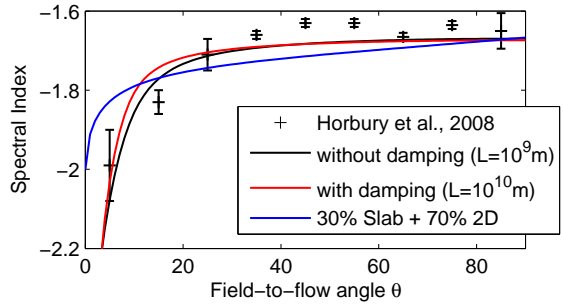}
 \caption{Spectral anisotropy found by \citet{Horbury2008} (black crosses), as obtained from our model with an outer scale of $L=10^9\,$m for undamped (black line) and $L=10^{10}\,$m for damped (red line) critical balance, as well as for slab+2-D turbulence with $30\%$ slab (blue line). \label{fig:horb}}
\end{figure}

\citet{Horbury2008} used magnetic field data from the Ulysses spacecraft at $1.4\,$AU in the fast solar wind ($V=750\,$km$\,$s$^{-1}$). From \citet{McComas2000}, we estimate the proton gyro radius as $\rho_i=190\,$km and due to lack of a better estimate, we assume $T_i=T_e$ to calculate the electron gyro radius. For comparison, we also calculate the best fit for a slab+2-D turbulence model, where we use a spectral index of $2$ for the slab component and $5/3$ for the 2-D component \citep{Horbury2011}. To quantify the goodness of fit we use a reduced error
\begin{equation}
	\chi = \sqrt{\frac{1}{N}\sum\limits_{i=1}^N\frac{(\bar \kappa_i-\kappa_i)^2}{\bar \sigma_i^2}} \ ,
\end{equation}
where $\bar \kappa_i$ and $\bar \sigma_i$ are the spectral indices and their corresponding errors, respectively, taken from \citet{Horbury2008}, $N$ is the number of angle bins, and $\kappa_i$ the modeled spectral indices, which are obtained by a least squares fit in the same frequency range, $15{-}98\,$mHz, as those used by \citet{Horbury2008}. The spectral anisotropies in the frequency range $15{-}98\,$mHz for the best fit parameters are shown in Figure \ref{fig:horb}.

In general, the fit is better for the critically balanced turbulence model ($\chi=2.8{-}3.0$) than for the slab+2-D turbulence model ($\chi=4.6$). This can be seen in Figure \ref{fig:horb}, where we show the observed and modeled spectral anisotropies. We find that an outer scale of $L=10^9\,$m in the undamped case and $L=10^{10}\,$m for the damped cascade give the best results. These values are in accordance with observations of the transition from $f^{-1}$-spectra to $f^{-5/3}$-spectra in the solar wind, which is believed to mark the end of the energy injection scale \citep{Schekochihin2009}. The spectra at small angles $\theta$ are found to be steeper than $\kappa=2$. This is not caused by damping. The difference between damped and undamped cascade is, according to Figures \ref{fig:kspa} and \ref{fig:kspa_damp}, very small. Instead, the steep slopes result from the fact that the fitting range $f=15-98\,$mHz includes the steeper kinetic range cascade, where slopes are significantly steeper than $\kappa=2$. For $\theta=5^\circ$, e.g., $f=98\,$mHz corresponds to $k_\perp\rho_i=1.8$.

For slab+2-D turbulence, we find the best fit has $30\%$ slab and $70\%$ 2-D turbulence, which is also in accordance with results obtained in the solar wind \citep{Bieber1996}. Note, however, that we use two different spectral indices for slab and 2-D turbulence while \citet{Bieber1996} used the same slope for both slab and 2-D turbulence.

\begin{figure}[t]
 \plotone{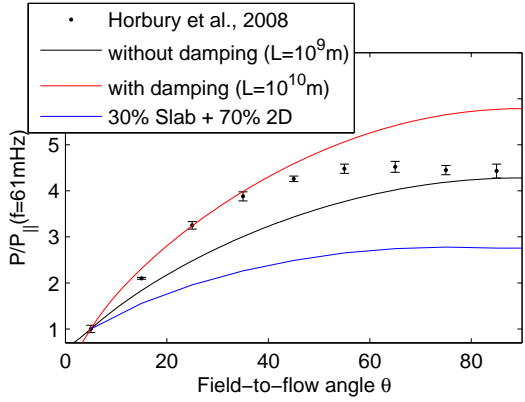}
 \caption{Power spectral anisotropy $P(\theta)/P(\theta{=}5^\circ)$ at frequency $f=61\,$mHz. The results of damped and undamped critical balance are close to the results of \citet{Horbury2008} while a combined model of slab and 2-D turbulence underestimates the anisotropy. \label{fig:horbani}}
\end{figure}

The anisotropy of the power, $P(\theta)/P(\theta{=}5^\circ)$, measured at a fixed frequency of $f=61\,$mHz is shown in Figure \ref{fig:horbani}. The results of the damped critically balanced cascade well reproduce the anisotropy found by \citet{Horbury2008} for angles $\theta<40^\circ$ but overestimate the power at larger angles. The undamped cascade generally shows a weaker anisotropy but is in agreement with the measurements at $\theta\sim90^\circ$. Similar to what has been found for the spectral index, a critically balanced cascade fits the data much better than slab+2-D turbulence.

Given the simplicity of our parameters, the presented fits to the results of \citet{Horbury2008} can be regarded as a qualitatively successful reproduction of the observed data. Although \citet{Forman2011} showed that the observed spectral anisotropy is in agreement with a critical balance on MHD scales, this is the first time that these results have been analyzed with a model including kinetic range scalings and damping terms. The quality of the fits indicates that the spectral anisotropy observed by \citet{Horbury2008} is consistent with a critically balanced cascade.

\subsection{Kinetic Range Turbulence}\label{sec:chen}
For an analysis of kinetic range magnetic field fluctuations, \citet{Chen2010a} used data from CLUSTER during fast solar wind conditions with moderate {ion} plasma $\beta_i\sim1$. They calculated the spectral index using a wavelet based method described by \citet{Horbury2008}. Ion and electron gyro radii as well as the bulk plasma velocity are given in Table 1 of \citet{Chen2010a} and the power anisotropy is given as $P_\perp/P_\|\sim20$. The spectral anisotropy is calculated separately for parallel and perpendicular fluctuations. Here, we limit our fit to the perpendicular fluctuations $P_\perp$. The spectral anisotropy of parallel fluctuations can not be reproduced sufficiently well, which is consistent with findings of \citet{Forman2013}. For the modeled spectra, we choose for the outer scales $L=10^9\,$m for the undamped and $L=10^{10}\,$m for the damped cascade based on our fit to the \citet{Horbury2008} data.

\begin{figure}[t]
 \plotone{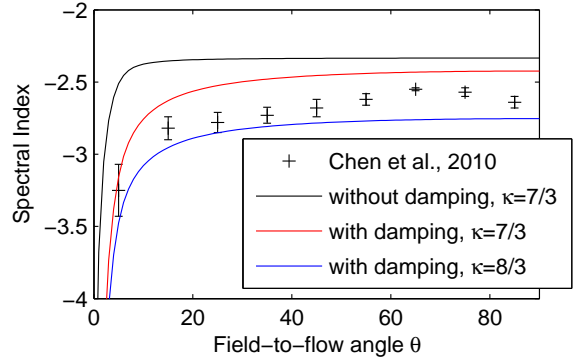}
 \caption{Spectral anisotropy of fluctuations perpendicular to the local magnetic field derived by \citet{Chen2010a} (black crosses) for kinetic scales $k\rho_i=[1.5,6]$ and numerically determined results for critically balanced KAW cascades. The spectral anisotropies for the damped cascades (red line for $\kappa=7/3$ and blue line for $\kappa=8/3$) fit better to the measured data compared to the undamped cascade (black line). \label{fig:chen}}
\end{figure}

In Figure \ref{fig:chen}, we show the results obtained from our forward calculation and those presented in \citet{Chen2010a} {for the ion kinetic range $1.5\leq k\rho_i\leq 6$}. Additional to the damped KAW cascade with $\kappa=7/3$, we show results for a damped critically balanced cascade with an assumed spectral index of $8/3$ and critical balance exponent $\alpha=1/3$. This scenario is included as hypothetical case for visual orientation only. Although the modeled results for damped and undamped spectra do not fit the data within error tolerances, the damped cascades show qualitatively similar spectral anisotropies. Here, the variation from $\theta=0^\circ$ to $\theta=90^\circ$ is smooth and not as abrupt as in the undamped case. For cross-comparison, we point the reader to Figure~\ref{fig:psd_theta_damp}, where it can be seen that PSD are steeper for smaller field-to-flow angles $\theta$. The undamped cascade, in contrast, leads to spectral slopes of ${\sim}7/3$ at almost all angles (cf.\ Figure~\ref{fig:psd_theta}). Even at $\theta=5^\circ$, where the observed spectral index is ${\sim}3.3$, the spectral index of the undamped cascade is already $2.4$. This is the consequence of the transition toward a quasi-perpendicular slope as was elaborated in section~\ref{sec:2d}. The spectral index of the damped KAW cascade on the other hand is $3.2$ at $\theta=5^\circ$, and thus much closer to the observation.

The results in Figure \ref{fig:chen} show that an undamped cascade leads to a sharp increase of the spectral index at small field-to-flow angles, which appears rather step-like. This result has interesting implications for the interpretation of critically balanced plasma turbulence. It is often assumed that a critically balanced cascade {in the ion kinetic range} can lead to any spectral slope, $7/3\leq\kappa\leq5$, for intermediate field-to-flow angles $\theta$. However, we have shown in Section \ref{sec:2d} and in equation~(\ref{eq:fmax}) that the spectral index is ${\sim}7/3$ for most angles. A smooth or slow variation of the spectral index, as seen in the results of \citet{Chen2010a}, can therefore not be caused by critical balance alone. Instead, the effect of damping is essential to obtain steeper spectra at non-zero angles, which means that the observed spectral anisotropy is to a large degree determined by the damping mechanisms.

\section{Modeling Spectral Densities at Saturn}\label{sec:satmod}
Recently, it has been shown that magnetic fluctuations in Saturn's plasma sheet form a turbulent cascade \citep{vonPapen2014b}. {These fluctuations have different properties compared to the solar wind. They are embedded within Saturn's strong background magnetic field and propagate as Alfv\'en waves along magnetic field lines until they are reflected by density gradients at the plasma sheet or the ionosphere.} Measurements by the Cassini spacecraft during its first seven orbits around Saturn indicate that observed power-law spectra can be interpreted as a critically balanced cascade of KAW. On MHD scales, however, power spectra indicating an Alfv\'en wave cascade with Kolmogorov-like power-law are observed only sporadically {because large-scale magnetospheric processes dominate this range of scales}. As Taylor's hypothesis holds in Saturn's magnetosphere for kinetic Alfv\'en waves with $k_\perp\gg k_\|$, we are able to test the observations with our forward model and to investigate the spatial distribution of observed spectral indices {as a function of radial distance to Saturn}.

As the turbulent cascade is predominantly observed on kinetic scales, we restrict our model to the reproduction of kinetic range spectra. We analyze the radial distribution of spectral indices reported in Figure 12 of \citet{vonPapen2014b} and focus on an explanation for the change of spectral slopes inside $9\,R_\mathrm{s}$, where $1\,R_\mathrm{s}=60268\,$km is the planetary radius of Saturn. Inside $9\,R_\mathrm{s}$, increasingly flatter spectra are observed, which so far could not be explained. As there are two distinct electron populations in Saturn's magnetosphere, we also analyze which dissipation scales control the onset of damping. Comparison of modeled and observed spectra may thus shed more light on the physics of the turbulence and may help to understand which electron population is energized by the turbulent cascade in Saturn's magnetosphere.

To model synthetic spectra in Saturn's magnetosphere, we use the parameters listed in Table~\ref{tab:para} as reported in \citet{vonPapen2014b}, where we use $2H_W\sim L$ as outer scale. These parameters fluctuate in time and vary strongly with radial distance to Saturn. The basic plasma parameters at Saturn - namely velocity, scale height, ion temperature and density - are based on observations by \citet{Thomsen2010} for water group ions, {which are the main constituent of Saturn's magnetosphere}. To calculate the electron gyro radii, which control the onset of the empirical damping term in equation~(\ref{eq:damp}), we use the electron temperature models given by \citet[][Table 1]{Schippers2008} derived from CAPS/MIMI measurements \citep{Young2004,Krimigis2004}. The field-to-flow angle $\theta$ is calculated assuming the plasma flow to be in azimuthal direction only. We further use the measured power anisotropy $\left<\frac{P_\perp}{P_\|}\right>$, averaged in the kinetic range, to estimate the ratio $\Psi/\Phi$ between the energy of toroidal and poloidal fluctuations according to equation~(\ref{eq:TP}). The slopes of the modeled spectra are calculated from the trace of the synthetic spectral tensor.

\begin{table}[ht]
  \begin{center}
  \caption{List of model input parameters to calculate synthetic spectra in Saturn's magnetosphere. \label{tab:para}}
  \begin{tabular}{cc}\tableline\tableline
    Parameter & Description \\ \tableline % & Source
    $\mathbf{B}$ & Magnetic field \\ % & measured \\
    $\theta$ & Field-to-flow angle \\ % & measured, \citet{Thomsen2010} \\
    $v$ & Plasma velocity \\ % & \citet{Thomsen2010} \\
    $H_W$ & Scale height of water group ions\\ % & \citet{Thomsen2010} \\
    $\rho_W$ & Water group ion gyro radius \\ % & measured, \citet{Thomsen2010} \\
    $\rho_e$ & Electron gyro radius \\ % & measured, \citet{Schippers2008} \\
    $V_\mathrm{A}$ & Alfv\'en velocity \\ % & measured, \citet{Thomsen2010} \\
    $\left<\frac{P_\perp}{P_\|}\right>$ & Power anisotropy \\ \tableline % & measured \\ \hline
  \end{tabular}
%   \tablecomments{We estimate the outer scale as twice the scale height of water group ions in the plasma sheet and the ratio of toroidal to poloidal fluctuations from the measured power anisotropy.}
  \end{center}
\end{table}

\begin{figure}[t]
 \plotone{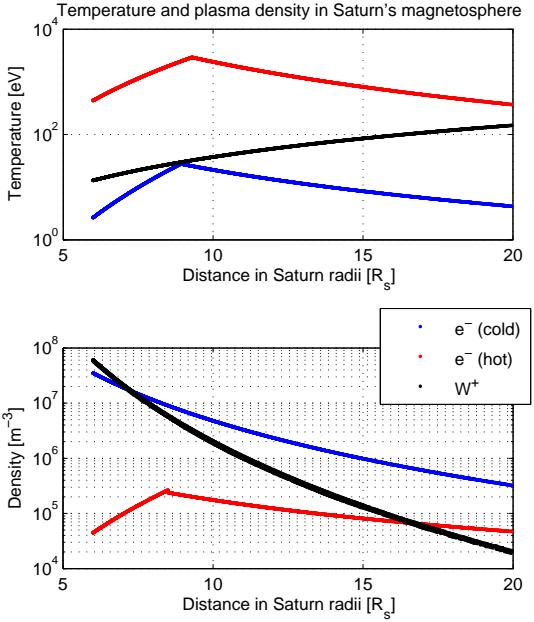}
 \caption{Equatorial electron temperatures (top) and densities (bottom) in Saturn's magnetosphere for the hot (red) and cold (blue) electron populations after \citet{Schippers2008}. For comparison, we also show the equatorial temperatures and densities of water group ions (black) as obtained from \citet{Thomsen2010}. \label{fig:Te}}
\end{figure}

Equatorial temperatures have been derived by \citet{Schippers2008} for the cold or thermal (${<}100\,$eV) and hot (${>}100\,$eV) electron populations. These temperatures control the electron gyro radii according to
\begin{equation}
	\rho_{e,c/h}=\frac{\sqrt{2 m_e k_B T_{e,c/h}}}{e B} \ ,
\end{equation}
where $T_{ec}$ and $T_{eh}$ denote the cold and hot electron temperatures, respectively. The radial profiles of the temperatures and densities are shown in the {top and bottom} panels of Figure~\ref{fig:Te}, respectively, together with the water group ion temperatures and densities for comparison. The model for the hot electrons is valid for distances up to $18\,R_\mathrm{s}$. The temperature of the cold population is more variable outside $15\,R_\mathrm{s}$ and thus only provided inside of that distance. However, due to the lack of a better estimate, we approximate the cold electron populations also outside of $15\,R_\mathrm{s}$ with this model. Both temperatures similarly peak at around $9\,R_\mathrm{s}$.

In the following, we model reduced spectra using electron gyro radii from the cold and hot electron populations and compare both to the observed spectra. The larger electron gyro radius of the hot electron population leads to an earlier onset of damping, i.e., the damping term is larger and its effects are already important at lower frequencies compared to damping on cold electron scales. This causes the slope of the hot electron spectrum to be slightly steeper than the cold electron spectrum.

In order to compare our modeled results with the observations, we need to add noise to the synthetic data and assume that the excitation of turbulence is not strong, i.e., we assume $\delta v_L < V_\mathrm{A}$. The latter is as expected from turbulence excited within a strong planetary magnetic field \citep{vonPapen2014b}. The noise levels are computed as the sum of magnetometer {instrument} noise $n_i$, quantization noise $n_q$, and aliasing. The instrument noise is measured and must therefore be added to the spectrum before calculating the aliasing. This leads to a synthetic spectrum $P'$ given by
\begin{eqnarray}
	P'(f) &=& P(f)/c_E + {n_\mathrm{i}/f} + {n_\mathrm{q}} \nonumber \\
	      & & + A(f)\cdot[ P(f)/c_E + {n_\mathrm{i}/f} ] \label{eq:pmod} \ ,
\end{eqnarray}
where $P(f)$ is the synthetic spectrum without noise, $c_E$ is a factor, which considers the weaker excitation discussed above, and the aliasing {function}
\begin{equation}
   A(f) = \sum\limits_{n=1}^\infty \frac{(f/f_\mathrm{ny})^\kappa}{[2n-(f/f_\mathrm{ny})]^\kappa} + \sum\limits_{n=1}^\infty \frac{(f/f_\mathrm{ny})^\kappa}{[2n+(f/f_\mathrm{ny})]^\kappa}
\end{equation}
is determined according to \citet{Podesta2006a} for a power-law of $\kappa=7/3$ {and Nyquist frequency $f_\mathrm{ny}$. In our case, this function takes on a value of $1.3$ at $f=f_\mathrm{ny}=3.6\,$Hz and drops rapidly for lower frequencies.}

The second term on the right-hand-side of equation~(\ref{eq:pmod}) is the instrument noise, which is $n_i=75\,$pT$^2/$Hz according to \citet{Dougherty2004} and goes with $1/f$ \citep{Russell1972}. The third term is the quantization noise, which can be estimated as $n_\mathrm{q}=\frac{1}{2}\Delta B^2 \Delta t$ \citep{Russell1972} with sampling period $\Delta t=0.14\,$s and the quantization of the magnetic field data, e.g., $\Delta B=4.9\,$pT in FGM range 0 \citep{Dougherty2004}. The factor $c_E$ {effectively corrects the energies of the synthetic spectra to the observed ones, however, it does not change the spectral index. The factor $c_E=\left<P_\mathrm{obs}/P\right>$} is on average $50$ and is calculated {as the geometrical mean of the ratio} of observed $P_\mathrm{obs}$ to synthetic spectra $P$ in the normalized wave number range $k_\perp\rho_W=[2,50]${, where $k_\perp=\frac{2\pi f}{V\sin(\theta)}$ and $\rho_W$ is the water group ion gyro radius,} and for signal-to-noise ratios $\mathrm{SNR}>5$.

\begin{figure}[t]
 \plotone{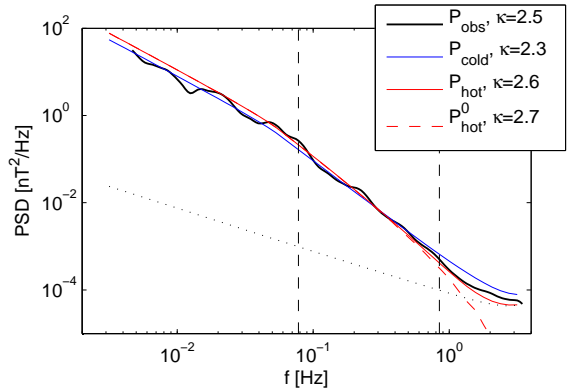}
 \caption{Power spectral density of a $10\,$min time series measured in Saturn's magnetosphere at a radial distance of $9.5\,R_\mathrm{s}$ (black line). The blue and red lines show synthetic spectra for damping controlled by the gyro radius of cold and hot electrons, respectively. Associated PSD are labeled $P_\mathrm{cold}$ and $P_\mathrm{hot}$, respectively. Vertical dashed lines show the fitting range in which the spectral index is calculated. The determined values are given in the legend. The dotted line shows the noise level corresponding to $P_\mathrm{hot}$ and the dashed red line the spectrum $P^0_\mathrm{hot}$ without noise. \label{fig:10msyn}}
\end{figure}

We carry out a forward modeling for each of the $1136$ observed $10\,$min time series and compute the spectral slopes in the same frequency ranges as determined for the observed data, i.e., in the range $2<k_\perp\rho_W<50$, and for $\mathrm{SNR}>5$ of the respective measurement \citep{vonPapen2014b}. As an example, we show an observed spectrum from $9.5\,R_\mathrm{s}$ and its corresponding synthetic spectra in Figure~\ref{fig:10msyn}. The black line shows the observed spectrum, the blue line the spectrum with dissipation controlled by the gyro radius of the cold electrons and the red line the spectrum with dissipation controlled by the gyro radius of the hot electrons. {All of these spectra include noise with exception of the red dashed line, which shows the spectrum $P^0_\mathrm{hot}$ without noise for comparison.}

Inside the fitting range, depicted by two vertical dashed lines, the three spectra are very similar. Their spectral indices are given in the legend of the figure and the modeled results ($\kappa_\mathrm{cold}=2.32$ and $\kappa_\mathrm{hot}=2.61$) are close to the observed slope of $\kappa=2.46$. However, at high frequencies the spectrum controlled by the gyro radius of the hot electrons clearly fits the observation better. At frequencies higher than the fitting range, the noise level shown by a dotted line leads to a flattening of the synthetic spectra. On average, this leads to a slight decrease of the spectral index within the fitting range by $0.06$ and $0.07$ for cold and hot electron spectra, respectively, compared to spectra without noise (see red dashed line in Figure \ref{fig:10msyn}).

\begin{figure}[t]
 \plotone{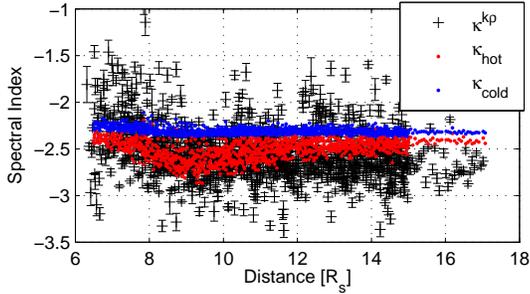}
 \caption{Radial distribution of observed spectral indices in Saturn's magnetosphere shown as black crosses. Spectral indices of synthetic spectra are shown as dots for damping on cold (blue) and hot (red) electron scales. Damping on hot electron scales explains qualitatively the change of slopes inside of $9\,R_\mathrm{s}$. \label{fig:kr_mod}}
\end{figure}

Figure \ref{fig:kr_mod} shows the radial distribution of observed spectral indices in Saturn's magnetosphere as black crosses (see Figure 12 in \citet{vonPapen2014b}). The spectral slopes for damping on cold electron scales (blue dots) are generally less steep than the observations and close to the undamped spectral index of $7/3$. This shows that the corresponding synthetic spectra have not yet reached dissipation scales, where the spectral energy decreases significantly. Damping on hot electron scales, on the other hand, leads to steeper spectra that agree much better with the observations. Clearly, the decrease of the electron temperature inside $9\,R_\mathrm{s}$ leads to shallower spectra because the damping is reduced for smaller electron gyro radii. This indicates that the change of slopes inside $9\,R_\mathrm{s}$ can be explained by damping effects. However, the radial profiles of both electron temperatures are nearly identical so that the difference between the two gyro radii can in principle be compensated by a simple factor $c>1$ in the exponent of the damping term, i.e., $\exp(-c\cdot k_\perp\rho_{ec})$ in equation~(\ref{eq:damp}). Note, that for larger distances ($>12\,R_\mathrm{s}$) the spectral indices $\kappa_\mathrm{hot}$ become shallower than the observations, which indicates either the presence of additional effects on the turbulent cascade or magnetospheric variations.

In summary, the forward modeling of kinetic range spectra has provided a possible explanation for the change of spectral slopes inside $9\,R_\mathrm{s}$, which is solely based on first principles of a KAW cascade and an empirical term to describe damping on electron scales. Our results indicate that the energy transferred along the kinetic range cascade is preferably deposited into the hot electron population. The heating examined here could thus be the primary process, which maintains the hot electron population in Saturn's magnetosphere.

\section{Conclusion}
We present results from a numerical {forward} model to evaluate reduced PSD from from an arbitrarily distributed energy density in three-dimensional $\mathbf{k}$-space\footnote{The forward code is available in the supplementary online material and under \url{www.geomet.uni-koeln.de/en/research/turbulence}.}. Given a critically balanced $\mathbf{k}$-space distribution of energy, we investigate the functional dependence of the {reduced} spectra on several parameters, such as the field-to-flow angle or the outer scale. Such an analysis has been carried out for the first time covering the complete range from MHD to electron kinetic scales. We show that for intermediate field-to-flow angles $\theta$, the spectral slope of undamped critically balanced turbulence is not constant, i.e., the reduced PSD is not an exact power-law (see Figure~\ref{fig:psd_theta}). Instead, the reduced spectra evolve with frequency in the undamped case toward a quasi-perpendicular spectrum. This is a pure sampling effect of highly Doppler-shifted measurements at any angle $\theta>0^\circ$ caused by the anisotropy $k_\perp\gg k_\|$ increasing with frequency. Power spectra in this quasi-perpendicular cascade range have a spectral slope corresponding to the spectral index of the perpendicular cascade $\kappa(\theta{=}90^\circ)$. For PSD that are additionally subject to damping (Figure \ref{fig:psd_theta_damp}), this transition to quasi-perpendicular spectra is masked by Landau damping on electron kinetic scales. Here, the spectra become steeper and steeper at higher frequencies.

The transition frequency, where the change of slope to $\kappa\approx\kappa(\theta{=}90^\circ)$ is reached for the undamped cascade, can be approximated by a simple expression $f_{2D}^\mathrm{a}$ given in equation~(\ref{eq:fmax}). Under typical solar wind conditions, $f_{2D}^\mathrm{a}$ for the {ion} kinetic regime is smaller than or on the same order of the observed spectral break at $f_b\sim0.3\,$Hz. Therefore, significantly steeper slopes than the perpendicular spectra on kinetic scales can only be explained by additional damping effects. We find that an empirical damping term at electron scales of the form $\exp(-k_\perp \rho_e)$ \citep{Alexandrova2012} already affects the spectral index in the ion kinetic range $\rho_i^{-1}<k<\rho_e^{-1}$ of the reduced spectrum and steepens it measurably.

We apply our model to in-situ measurements in the solar wind and show that turbulent fluctuations measured by \citet{Horbury2008} are in good agreement with a critically balanced cascade and less so with a combined slab+2-D turbulence. In the kinetic range we find good agreement with the results of \citet{Chen2010a}, which can be explained by a damped critically balanced KAW cascade. While the spectral break and damping terms have only minor influence on the fit to the observations of \citet{Horbury2008}, the inclusion of a damping term is essential to explain the results in the kinetic range. Indeed, spectral slopes for an undamped cascade differ strongly from the \citet{Chen2010a} results and are approximately $7/3$ for nearly all field-to-flow angles {in the ion kinetic regime}. This is caused by the transition toward a quasi-perpendicular slope. This means that damping is an integral part of the kinetic range cascade and the dominating factor in the observed spectral anisotropy.

The forward modeling technique is further applied to observations of turbulent magnetic field fluctuations in Saturn's magnetosphere. Here, we show that the measured reduced spectra are qualitatively in agreement with a critically balanced kinetic Alfv\'en wave cascade, which further corroborates the interpretation of \citet{vonPapen2014b}. The observation of shallower spectra inside $9\,R_\mathrm{s}$ can be reproduced by damping on electron scales controlled by the hot electron population. This indicates that the dissipation of turbulent magnetic fluctuations predominantly heats the hot electron population {at Saturn}. This additional insight on magnetic turbulence in Saturn's magnetosphere shows the advantage of applying a {first principles} forward modeling technique {to determine the shape of reduced spectra}.

%% If you wish to include an acknowledgments section in your paper,
%% separate it off from the body of the text using the \acknowledgments
%% command.

%% Included in this acknowledgments section are examples of the
%% AASTeX hypertext markup commands. Use \url without the optional [HREF]
%% argument when you want to print the url directly in the text. Otherwise,
%% use either \url or \anchor, with the HREF as the first argument and the
%% text to be printed in the second.

\acknowledgments
{We like to thank R.\ A.\ Burger for pointing out the relevance of our study to cosmic ray transport and the referee for constructive comments. The authors also acknowledge fruitful discussions with R.\ Wicks and M.\ Forman.}

\appendix

\section{Estimation of Transition Frequency to Quasi-Perpendicular Spectra}\label{app}
Here, we derive an approximate expression for the frequency $f_{2D}^\mathrm{a}$, at which the reduced power spectrum of an undamped critically balanced cascade turns quasi-perpendicular, i.e., the PSD is characterized by a spectral slope $\kappa=\kappa(\theta{=}90^\circ)+\Delta\kappa$ with a small error $\Delta\kappa$. For mathematical tractability we derive this approximate expression under additional assumptions. {Later}, we show by comparison with our full three-dimensional model that the resultant expression still provides a good estimator when these assumptions are relaxed. The assumptions we use for our derivation are a two-dimensional $\mathbf{k}$-space ($k_\|$, $k_\perp$) and a critically balanced cascade that is controlled by a Dirac Delta function instead of the exponential function in equation~(\ref{eq:emhd}). The latter seems to be a strong simplification but \citet{Forman2011} and \citet{Turner2012b} have shown that there are only minor differences between the spectral anisotropies of PSD controlled by an exponential and a Dirac Delta function.

In order to estimate the implications of our assumption of a two-dimensional $\mathbf{k}$-space, it is important to clarify the three-dimensional geometry of the problem. In Figure \ref{fig:kspa}, we see a slice along the $k_x$-$k_z$ plane. While the planes of integration {(white dashed lines)} are two-dimensional and expand straight in $k_y$ direction, the energy distribution is three-dimensional and axisymmetric with respect to $k_z$. This means that the plane of integration intersects with both Dirac Delta surfaces (for $k_z>0$ and $k_z<0$) in a complicated curve, which depends on outer scale $L$, field-to-flow angle $\theta$, and critical balance exponent $\alpha$. However, most of the energy along these curves resides at $k_y=0$ and decreases with growing wave number in $y$-direction. Therefore, we assume that the assumption of a two-dimensional wave vector space is suitable for our following derivation. This is subsequently corroborated by the agreement of the such derived expression with our fully three-dimensional model results.

\begin{figure}[t]
 \epsscale{.50}\plotone{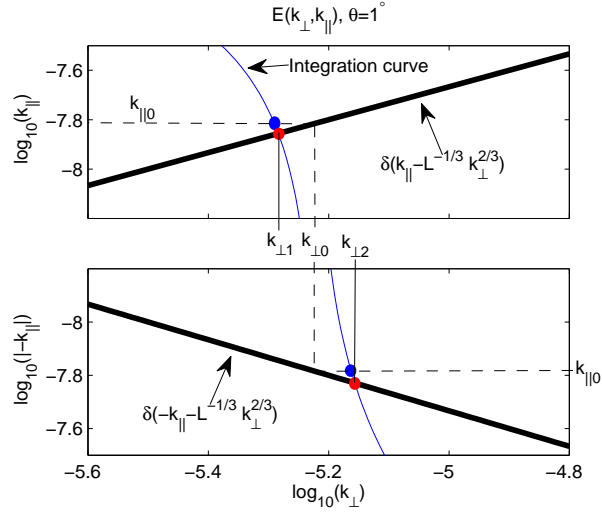}
 \caption{Double-logarithmic plot of energy distribution given by a Delta correlated critical balance model as given in equation~(\ref{eq:deltaE}) (thick black line) in $\mathbf{k}$-space with blue line showing the integration curve for a field-to-flow angle $\theta=1^\circ$. Red circles indicate intersections of integration curve with Dirac Delta {branch} and thin black lines denote the locations of $k_{\perp,1/2}$. The blue circles show the locations approximated according to equation~(\ref{eq:kp12prox}). Dashed lines show the locations of $k_{\perp0}$ and $k_{\|0}$. \label{fig:kp12}}
\end{figure}

In two dimensions, the energy distribution of critical balance controlled by a Dirac Delta function may be written as
\begin{equation}
  E_{2D}(k_\perp,k_\|) \propto k_\perp^{-\kappa} \, \delta\!\left(|k_\||-L^{-1/3}\rho^{\alpha-2/3} k_\perp^\alpha\right) \ , \label{eq:deltaE}
\end{equation}
where $\kappa=[5/3, 7/3, 8/3]$ are the one-dimensional spectral indices and $\alpha=[2/3, 1/3, 0]$ the critical balance exponents on MHD, {KAW, and ED} scales, respectively \citep[see also][]{Grappin2010}. {For the sake of simplicity, we use only one kinetic scale $\rho$ in equation (\ref{eq:deltaE}): in the KAW range $\rho=\rho_i$, while the result for the ED range can be obtained by setting $\rho=\sqrt{\rho_i\rho_e}$. The distribution of energy density in $\mathbf{k}$-space given by $E_{2D}$ is shown in Figure \ref{fig:kp12}.} The intersection of {the two Dirac Delta branches} with the line of integration given by equation~(\ref{eq:plane}) are two points:
\begin{equation}
  {\pm}k_\| = L^{-1/3}\rho^{\alpha-2/3} k_\perp^\alpha \ , \label{eq:pmkpa}
\end{equation}
positive $k_{\|,1}(k_{\perp,1})$ and negative $k_{\|,2}(k_{\perp,2})$. The locations of these points are depicted as red circles in Figure \ref{fig:kp12}. Field-to-flow angles smaller than $90^\circ$ yield $k_{\perp,1}<k_{\perp,0}<k_{\perp,2}$, where
\begin{equation}
  k_{\perp,0}=\frac{2\pi f} {v \sin(\theta)} \label{eq:kpo}
\end{equation}
is the wave number corresponding to $k_\|=0$. Note, that we do not need to make any assumptions regarding the energy at zero parallel wave number. The two locations $k_{\perp,1/2}$ are uniquely defined by equations (\ref{eq:plane}) and (\ref{eq:pmkpa}). It follows that
\begin{eqnarray}
	k_{\perp,1/2}&=&\frac{2\pi f}{v\sin(\theta)}\mp k_{\|,1/2} \cot(\theta) \nonumber \\
		     &=&\frac{2\pi f}{v\sin(\theta)}\mp L^{-1/3}\rho^{\alpha-2/3}k_{\perp,1/2}^\alpha \cot(\theta) \label{eq:kp12} \ .
\end{eqnarray}
With the particular choices of $\alpha=[2/3, 1/3]$, this equation can be written as a cubic polynomial with real solutions $k_{\perp,1/2}(\theta)$. Even though a cubic algebraic equation can be solved analytically, we search for an approximate solution to achieve a mathematically simpler expression, which is more easily to work with.

Let us now look at the energy $E_{2D}$ in $\mathbf{k}$-space that contributes to the integration:
\begin{equation}
  E_{2D}\propto k_{\perp,1}^{-\kappa}+k_{\perp,2}^{-\kappa} \ ,
\end{equation}
which is essentially the summation over both critical balance branches. In the following, we show that this energy asymptotically approaches $E_0\propto 2k_{\perp,0}^{-\kappa}$ with increasing frequency, thus leading to a scaling according to $\theta=90^\circ$. At a certain frequency $f_{2D}^\mathrm{a}$, the measurement uncertainty will be larger than the difference between the energies $E_{2D}$ and $E_0$ and, therefore, the observed PSD will scale like the perpendicular cascade. We proceed to estimate the frequency for which the reduced PSD turns quasi-perpendicular by demanding that the ratio $E_{2D}/E_0$ be almost unity. Because the spectral index is obtained from the logarithms of the power spectral energy, we demand
\begin{equation}
	\ln\left[\frac{k_{\perp,1}^{-\kappa}+k_{\perp,2}^{-\kappa}}{2k_{\perp,0}^{-\kappa}}\right] < \epsilon \label{eq:eps} \ ,
\end{equation}
where $\epsilon>0$ is chosen according to typical measurement uncertainties $\Delta\kappa$ for spectral slopes in solar wind observations.

To proceed further, we approximate
\begin{equation}
  k_{\|,1/2} \approx k_{\|0} = L^{-1/3}\rho^{\alpha-2/3}k_{\perp,0}^\alpha \ . \label{eq:kp12prox}
\end{equation}
The geometrical interpretation of this approximation is shown in Figure \ref{fig:kp12}. Instead of the exact solutions shown by the red circles, we now evaluate the integral at the approximated locations shown by the blue circles. These are slightly shifted along the integration curve toward lower $k_\perp$-values. Now, we can write equation~(\ref{eq:kp12}) as
\begin{eqnarray}
	k_{\perp,1/2} &\approx& k_{\perp,0} \mp L^{-1/3}\rho^{\alpha-2/3} k_{\perp,0}^\alpha \cot(\theta) \nonumber \\
		      &=&	 k_{\perp,0} \mp \Delta \label{eq:kp0} \ .
\end{eqnarray}
Inserting this in equation~(\ref{eq:eps}), we get
\begin{equation}
	\left(1-\frac{\Delta}{k_{\perp,0}}\right)^{-\kappa}+\left(1+\frac{\Delta}{k_{\perp,0}}\right)^{-\kappa} < 2 e^\epsilon  \label{eq:eps2} \ .
\end{equation}
If we expand the two terms on the left-hand-side of equation~(\ref{eq:eps2}) to second order around $\Delta/k_{\perp,0}=0$, we find that
\begin{equation}
    \frac{\Delta}{k_{\perp,0}} < \sqrt{ \frac{2e^\epsilon-2}{\kappa(\kappa+1)} } \label{eq:eps3} \ .
\end{equation}
Inserting $\Delta=L^{-1/3}\rho^{\alpha-2/3} k_{\perp,0}^\alpha \cot(\theta)$ and $k_{\perp,0}=2\pi f/\left(v\sin(\theta)\right)$ into this equation finally yields
\begin{equation}
    f > \frac{v\sin(\theta)}{2\pi} L^{\frac{1}{3\alpha-3}} \rho^\frac{2-3\alpha}{3\alpha-3} \left(\frac{2e^\epsilon-2}{\kappa(\kappa+1)}\right)^\frac{1}{2\alpha-2} \cot(\theta)^\frac{1}{1-\alpha}\label{eq:fmax2} \ .
\end{equation}

For an appropriately chosen $\epsilon$, the difference between energies $E_{2D}$ and $E_0$ for all frequencies $f>f_{2D}^\mathrm{a}$ is so small that the scaling will be quasi-perpendicular within measurement errors $\Delta\kappa$. Note, that the change of slope in frequency space is a pure sampling effect and does not mean that the nature of the turbulent cascade in wave vector {space} is changing at this frequency. For $\epsilon\ll1$, we can further simplify equation~(\ref{eq:fmax2}) to determine the {approximate transition} frequency that marks the boundary to quasi-perpendicular scaling as
\begin{equation}
    f_{2D}^\mathrm{a} = \frac{v\sin(\theta)}{2\pi} \left(L^{\frac{1}{3}} \rho^\frac{2-3\alpha}{3} \sqrt{\frac{2\epsilon}{\kappa(\kappa+1)}} \tan(\theta) \right)^{\frac{1}{\alpha-1}} \label{eq:fmax} \ .
\end{equation}
{Note, that equation (\ref{eq:fmax}) can additionally be used to test whether an observed steep slope in a reduced spectrum of critically balanced turbulence can be explained by spectral anisotropy only, i.e., without damping, or if additional damping is needed to explain the steep slope.}

In order to check in which turbulent range $f_{2D}^\mathrm{a}$ lies, i.e., in the MHD, the KAW or the ED range, the following procedure might be used: One chooses the appropriate critical balance exponent for the observed range of scales, namely $\alpha=2/3$ for MHD, $\alpha=1/3$ for KAW, or $\alpha=0$ for electron kinetic scales. Inserting $\alpha$ in equation (\ref{eq:fmax}) together with the relative plasma velocity $v$, observation angle $\theta$, outer scale $L$, and controlling kinetic scale $\rho$ ($\rho=\rho_i$ for KAW and $\rho=\sqrt{\rho_i\rho_e}$ for ED), one obtains the transition frequency $f_{2D}^\mathrm{a}$ for the observation geometry under consideration. All frequencies $f>f_{2D}^\mathrm{a}$ will be characterized by a quasi-perpendicular spectrum.

To estimate in which range (MHD, KAW, ED) the transition toward quasi-perpendicular spectra occurs, one can calculate the corresponding perpendicular wave number according to $k_\perp= 2 \pi f_{2D}^\mathrm{a}/(v \sin(\theta))$. If the resultant $k_\perp$ falls outside the range used to calculate $f_{2D}^\mathrm{a}$ (with certain $\alpha$ and $\rho$), then the transition toward quasi-perpendicular spectra will either happen on smaller scales (if $k_\perp$ is too large for the corresponding range) or the transition has already happened on larger scales (if $k_\perp$ is too small). The latter case, e.g., is found for the ED range in the solar wind as the transition toward quasi-perpendicular spectra already occurs on KAW scales. To find the correct transition frequency, the procedure then needs to be repeated with the $\alpha$ and $\rho$ values from the alternative range.

Note, that the error parameter $\epsilon$ must be determined according to the measurement accuracy, which usually requires modeling the corresponding spectra. We find that $\epsilon=0.05$ in equation~(\ref{eq:fmax}) represents an uncertainty of the spectral index of $\Delta\kappa=\pm0.1$ in the KAW and $\Delta\kappa=\pm0.02$ in the MHD range {for typical solar wind conditions}. For the {KAW} range, this uncertainty is on the order of the measurement error in solar wind observations.

In the following, we test {the approximate expression $f_{2D}^\mathrm{a}$ given in} equation~(\ref{eq:fmax}) against our three-dimensional model results. {For that matter, we determine the transition frequencies $f_{2D}^\mathrm{mod}$ from modeled} PSD without a break at $k_\perp\rho_i\sim1$ or $k_\perp\rho_e\sim1$, i.e., for equations (\ref{eq:emhd}) and (\ref{eq:ekaw}) separately, each extending over the whole range of wave vectors. Because of divergence at low wave number $k_\perp$, this method is numerically not stable if we use an energy distribution in $\mathbf{k}$-space, which is solely given by equation~(\ref{eq:ed}). From the modeled PSD {in the MHD and KAW range}, we calculate the spectral slopes between two consecutive data points ($f_{n+1}/f_n\approx1.3$) as $\kappa=\kappa(\theta{=}90^\circ)+\Delta\kappa$, where $\kappa(\theta{=}90^\circ)=[5/3,7/3]$ in the MHD and {KAW} regime, respectively. {We then define $f_{2D}^\mathrm{mod}$ as the frequency, where $\Delta\kappa$ decreases below thresholds of $\Delta\kappa=\pm0.1$ in the KAW and $\Delta\kappa=\pm0.02$ in the MHD range.}

\begin{figure}[t]
 \epsscale{.50}\plotone{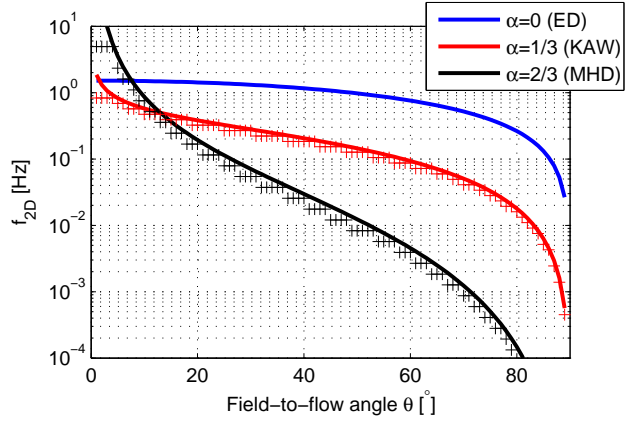}
 \caption{Transition frequencies as a function of $\theta$ for critical balance values $\alpha=[2/3, 1/3, 0]$. Solid lines show $f_{2D}^\mathrm{a}$ calculated from the approximate expression in equation~(\ref{eq:fmax}) {using $\epsilon=0.05$} and crosses show $f_{2D}^\mathrm{mod}$ obtained numerically from modeled PSD for an uncertainty in the spectral index of $\Delta\kappa=0.02$ on MHD and $\Delta\kappa=0.1$ on ion kinetic scales. \label{fig:fmax}}
\end{figure}

In Figure~\ref{fig:fmax}, we show the such derived $f_{2D}^\mathrm{mod}$ (crosses) for parameters given in Table \ref{tab:sw} compared to the analytical approximation $f_{2D}^\mathrm{a}$ (solid lines) given in equation~(\ref{eq:fmax}) for critical balance values $\alpha=[2/3,1/3,0]$ as a function of field-to-flow angle $0^\circ<\theta<90^\circ$. The {modeled} three-dimensional results agree well with the estimations from equation~(\ref{eq:fmax}) for $\epsilon=0.05$, which are shown by solid lines for the MHD, KAW, and ED range. On MHD scales, the transition toward quasi-perpendicular spectra occurs over a broad frequency range and is therefore not as apparent as on kinetic scales.

The calculated transition frequencies $f_{2D}^\mathrm{a}$ are shown in the corresponding figures throughout this paper as colored dots for the KAW range. The transition frequencies $f_{2D,\mathrm{MHD}}^a$ for the MHD regime according to equation~(\ref{eq:fmax}) are not shown in the figures as they usually fall outside the MHD range, i.e., $f_{2D,\mathrm{MHD}}^a>\frac{V}{2\pi\rho_i}$. However, for $\theta=20^\circ$ equation~(\ref{eq:fmax}) results in $f_{2D,\mathrm{MHD}}^a=0.2\,$Hz. The transition frequencies according to the ED range always appear at lower frequencies than the electron spectral break and are therefore not shown.

\nocite{Engelbrecht2014,Engelbrecht2015}

\clearpage

\end{document}